\documentclass[aps,onecolumn,10pt]{revtex4}
\usepackage{amsmath}
\usepackage{amssymb}
\usepackage{hyperref}
\hypersetup{colorlinks=true} 
\numberwithin{equation}{section}
\numberwithin{equation}{section}

\begin{document}
\allowdisplaybreaks
\setcounter{equation}{0}

\title{Exact solution to perturbative conformal cosmology from recombination until the current era}

\author{Asanka Amarasinghe, Tianye Liu, Daniel A. Norman and Philip D. Mannheim}
\affiliation{Department of Physics, University of Connecticut, Storrs, CT 06269, USA \\
asanka.amarasinghe@uconn.edu, tianye.liu@uconn.edu,daniel.norman@uconn.edu, philip.mannheim@uconn.edu\\ }

\date{January 5 2021}

\begin{abstract}

In a previous paper (P. D. Mannheim, Phys. Rev. D 102, 123535 (2020)) we studied cosmological perturbation theory in the cosmology associated with the fourth-order derivative conformal gravity theory, and provided an exact solution to the theory in the recombination era. In this paper we present an exact solution that holds all the way from recombination until the current era.

\end{abstract}

\maketitle

\section{Introduction}
\label{S1}

\subsection{Motivation}
\label{S1az}

A primary interest of cosmological research has been the study of cosmological fluctuations around a homogeneous and isotropic cosmic microwave background (see e.g.  \cite{Dodelson2003,Mukhanov2005,Weinberg2008,Lyth2009,Ellis2012}). The focus in the main has been on the Einstein-gravity-based inflationary universe model \cite{Guth1981}, which leads to a concordance picture (see e.g. \cite{Bahcall2000,deBernardis2000,Tegmark2004}) of a spatially flat universe composed primarily of dark matter and dark energy. Given the lack to date of the detection of any dark matter candidates, given the lack of understanding of the dark energy or cosmological  constant 60 orders of magnitude fine tuning problem, and given the presumption that the classical gravity treatment of the model that is made would not be destroyed by uncontrollably infinite quantum gravitational radiative corrections \cite{footnoteZ1}, some candidate alternative proposals have been advanced in the literature. In this paper we consider one specific alternative, namely conformal gravity 
(see \cite{Mannheim1990,Mannheim1989,Mannheim1994,Mannheim2006,Mannheim2012b,Mannheim2017} and references therein). As a candidate gravitational theory conformal gravity has been shown capable of eliminating the need for galactic dark matter not only by providing fits to a wide class of galactic rotation curves without any need for dark matter but by doing so with universal galaxy-independent parameters \cite{Mannheim2011b,Mannheim2012c,O'Brien2012}, even as, in contrast, one currently does have to introduce galaxy-dependent free parameters in dark matter fits. Also it was shown that through its underlying local conformal symmetry (invariance under $g_{\mu\nu}(x)\rightarrow e^{2\alpha(x)}g_{\mu\nu}(x)$ with a spacetime-dependent $\alpha(x)$) conformal gravity controls the cosmological constant without fine tuning. And with the conformal symmetry requiring that the gravitational sector coupling constant $\alpha_g$ be dimensionless, the conformal theory is renormalizable. With conformal gravity also being quantum-mechanically ghost free and unitary \cite{Bender2008a,Bender2008b,Mannheim2011a,Mannheim2018}, conformal gravity  provides a consistent quantum gravity theory in  four spacetime dimensions.  

From the perspective of conformal gravity the dark matter, dark energy and quantum gravity problems are not three separate problems at all. Rather, they all have a common cause, namely the extrapolation of Newton-Einstein gravity beyond its solar system origins. Consequently,  they can all have a common solution, with conformal gravity endeavoring to provide such a solution through a different extrapolation of solar system wisdom. However, even though the conformal theory has successfully been applied to the homogeneous and isotropic cosmological background by providing \cite{Mannheim1992,Mannheim2006,Mannheim2012b,Mannheim2017} a horizon-free background cosmology with no flatness problem, while providing a very good, non-fine-tuned, dark-matter-free  fit to the accelerating universe supernovae data of \cite{Riess1998,Perlmutter1999}, it still needs to be applied to the fluctuations around that background. Such a study is now in progress, with the initial development of the cosmological perturbation theory that is required having been presented in general in \cite{Mannheim2012a,Amarasinghe2019,Phelps2019,Mannheim2020,Amarasinghe2020}. In this paper we take a further step by  providing a new exact solution to the fourth-order derivative conformal gravity cosmological fluctuation equations that holds all the way from recombination until the current era. We had in fact already provided an exact solution that holds at recombination itself \cite{Mannheim2020}, and in this paper we build on that study. In regard to conformal gravity we note also that various other studies of conformal gravity and of higher derivative gravity theories in general can be found in \cite{Hoyle1964,Stelle1977,Stelle1978,Adler1982,Lee1982,Zee1983,Riegert1984a,Riegert1984b,Teyssandier1989,'tHooft2010a,'tHooft2010b,'tHooft2011,'tHooft2015a,Maldacena2011,Horne2016}.

In \cite{Mannheim2020} we derived the conformal gravity cosmological fluctuation equations in all generality, for any background cosmological geometry and any set of background matter sources. While these equations hold for any background expansion radius $a(t)$ and any background spatial three-curvature $k$, in \cite{Mannheim2020} we solved them exactly at recombination in the case of most interest to conformal cosmology, namely  negative $k$ \cite{Mannheim2006,Mannheim2012b,Mannheim2017}, a feature of the theory that  we elaborate on below. To actually study the gravitational fluctuation equations we used  the scalar, vector, tensor expansion of the fluctuation metric that was first introduced in  \cite{Lifshitz1946}  and \cite{Bardeen1980} and then widely applied in  perturbative cosmological studies  (see e.g. \cite{Kodama1984,Mukhanov1992,Stewart1990,Ma1995,Bertschinger1996,Zaldarriaga1998} and \cite{Dodelson2003,Mukhanov2005,Weinberg2008,Lyth2009,Ellis2012}). This expansion is based on quantities that transform as three-dimensional scalars, vectors and tensors, and as such it is particularly well suited to Robertson-Walker geometries because such geometries have a spatial sector that is maximally three-symmetric. Even though the scalar, vector, tensor expansion  is based on quantities whose transformation properties are defined with respect to three spatial dimensions rather than four spacetime dimensions, nonetheless using the scalar, vector, tensor basis leads to fluctuation equations that are composed of combinations of them that are fully four-dimensionally gauge invariant. They are thus very convenient for cosmological fluctuation theory. Thus in this paper we shall follow the procedure laid out in \cite{Mannheim2020}.

This paper is organized as follows. In Sec. \ref{S1} we describe the background gravitational and matter sectors. In Secs. \ref{S2} and \ref{S3} we consider fluctuations around the background using the scalar, vector, tensor fluctuation basis, and derive the cosmological fluctuation equations of interest to us here. Some of the material presented in these sections has already been presented in \cite{Mannheim2020}, but is included here for completeness, and so that it can be adapted for the purposes of this paper. In Sec. \ref{S4} we solve the fluctuation equations in the scalar sector in order to obtain the time behavior of the scalar modes, doing so for both conformal time and comoving time. In Secs. \ref{S5} and \ref{S6} we provide an analogous discussion for the vector modes and for the tensor modes. In Sec. \ref{S7} we present the spatial behavior of the scalar, vector and tensor modes, and give normalization factors for the modes. Finally, in Sec. \ref{S8} we collect together all of our results, and give the full temporal and spatial behavior of the scalar, vector and tensor modes all the way from recombination until the current era.

\subsection{The Background Conformal Gravity Cosmology -- Gravity Sector}
\label{S1bz}

Conformal gravity is a pure metric theory of gravity that possesses all of the general coordinate invariance and equivalence principle structure of standard gravity while augmenting it with an additional symmetry, local conformal invariance, in which  the action is left invariant under local conformal transformations on the metric of the form $g_{\mu\nu}(x)\rightarrow e^{2\alpha(x)}g_{\mu\nu}(x)$ with arbitrary local phase $\alpha(x)$. Under such a symmetry a gravitational action that is to be a polynomial function of the Riemann tensor is uniquely prescribed, and with use of the Gauss-Bonnet theorem is given by (see e.g. \cite{Mannheim2006}) 
\begin{eqnarray}
I_{\rm W}=-\alpha_g\int d^4x\, (-g)^{1/2}C_{\lambda\mu\nu\kappa}
C^{\lambda\mu\nu\kappa}
\equiv -2\alpha_g\int d^4x\, (-g)^{1/2}\left[R_{\mu\kappa}R^{\mu\kappa}-\frac{1}{3} (R^{\alpha}_{\phantom{\alpha}\alpha})^2\right].
\label{1.1z}
\end{eqnarray}
Here $\alpha_g$ is a dimensionless  gravitational coupling constant, and
\begin{eqnarray}
C_{\lambda\mu\nu\kappa}= R_{\lambda\mu\nu\kappa}
-\frac{1}{2}\left(g_{\lambda\nu}R_{\mu\kappa}-
g_{\lambda\kappa}R_{\mu\nu}-
g_{\mu\nu}R_{\lambda\kappa}+
g_{\mu\kappa}R_{\lambda\nu}\right)
+\frac{1}{6}R^{\alpha}_{\phantom{\alpha}\alpha}\left(
g_{\lambda\nu}g_{\mu\kappa}-
g_{\lambda\kappa}g_{\mu\nu}\right)
\label{1.2z}
\end{eqnarray}
is the conformal Weyl tensor. (Here and throughout we follow the notation and conventions of \cite{Weinberg1972}.)

With the Weyl action $I_{\rm W}$ given in  (\ref{1.1z}) being a fourth-order derivative function of the metric, functional variation with respect to the metric $g_{\mu\nu}(x)$ generates fourth-order derivative gravitational equations of motion of the form \cite{Mannheim2006} 
\begin{eqnarray}
-\frac{2}{(-g)^{1/2}}\frac{\delta I_{\rm W}}{\delta g_{\mu\nu}}=4\alpha_g W^{\mu\nu}=4\alpha_g\left[2\nabla_{\kappa}\nabla_{\lambda}C^{\mu\lambda\nu\kappa}-
R_{\kappa\lambda}C^{\mu\lambda\nu\kappa}\right]=4\alpha_g\left[W^{\mu
\nu}_{(2)}-\frac{1}{3}W^{\mu\nu}_{(1)}\right]=T^{\mu\nu},
\label{1.3z}
\end{eqnarray}
where the functions $W^{\mu \nu}_{(1)}$ and $W^{\mu \nu}_{(2)}$ (respectively associated with the $(R^{\alpha}_{\phantom{\alpha}\alpha})^2$ and $R_{\mu\kappa}R^{\mu\kappa}$ terms in (\ref{1.1z})) are given by
\begin{eqnarray}
W^{\mu \nu}_{(1)}&=&
2g^{\mu\nu}\nabla_{\beta}\nabla^{\beta}R^{\alpha}_{\phantom{\alpha}\alpha}                                             
-2\nabla^{\nu}\nabla^{\mu}R^{\alpha}_{\phantom{\alpha}\alpha}                          
-2 R^{\alpha}_{\phantom{\alpha}\alpha}R^{\mu\nu}                              
+\frac{1}{2}g^{\mu\nu}(R^{\alpha}_{\phantom{\alpha}\alpha})^2,
\nonumber\\
W^{\mu \nu}_{(2)}&=&
\frac{1}{2}g^{\mu\nu}\nabla_{\beta}\nabla^{\beta}R^{\alpha}_{\phantom{\alpha}\alpha}
+\nabla_{\beta}\nabla^{\beta}R^{\mu\nu}                    
 -\nabla_{\beta}\nabla^{\nu}R^{\mu\beta}                       
-\nabla_{\beta}\nabla^{\mu}R^{\nu \beta}                          
 - 2R^{\mu\beta}R^{\nu}_{\phantom{\nu}\beta}                                    
+\frac{1}{2}g^{\mu\nu}R_{\alpha\beta}R^{\alpha\beta},
\label{1.4z}
\end{eqnarray}                                 
and where $T^{\mu\nu}$ is the conformal invariant energy-momentum tensor associated with a conformal matter source \cite{footnoteZ2}.  Since $W^{\mu\nu}=W^{\mu\nu}_{(2)}-(1/3)W^{\mu\nu}_{(1)}$, known as the Bach tensor \cite{Bach1921},  is obtained from an action that is both general coordinate invariant and conformal invariant, in consequence, and without needing to impose any equation of motion or stationarity condition, $W^{\mu\nu}$ is automatically covariantly conserved and covariantly traceless and obeys $\nabla_{\nu}W^{\mu\nu}=0$, $g_{\mu\nu}W^{\mu\nu}=0$ on every variational path used for the functional variation of $I_{\rm W}$. With the Weyl tensor vanishing in geometries that are conformal to flat, in the conformal to flat Robertson-Walker and de Sitter geometries of interest to cosmology the background $W^{\mu\nu}$ is zero. Despite this,  fluctuations around that background are not conformal to flat, with the fluctuating $\delta W^{\mu\nu}$ then being nonzero. However, with $g_{\mu\nu}W^{\mu\nu}=0$ it follows that $\delta g_{\mu\nu}W^{\mu\nu}+g_{\mu\nu}\delta W^{\mu\nu}=0$. Then with the cosmological background obeying $W^{\mu\nu}=0$, it follows that $g_{\mu\nu}\delta W^{\mu\nu}=0$. Thus only nine of the ten components of the fluctuating $\delta W^{\mu\nu}$ are independent.

\subsection{The Background Cosmological Model}
\label{S1ez}

Since particles can only acquire mass in a conformal invariant theory by symmetry breaking, we introduce a scalar field $S(x)$ for this purpose. Even though it will not actually matter below since in the end the only thing we will need from the matter sector fields is a perfect fluid form for their energy-momentum tensor, for illustrative purposes we take the matter sector fields to be represented by fermions, with the conformally invariant matter sector action then being of the form  
\begin{eqnarray}
I_M=-\int d^4x(-g)^{1/2}\left[\frac{1}{2}\nabla_{\mu}S
\nabla^{\mu}S-\frac{1}{12}S^2R^\mu_{\phantom         
{\mu}\mu}
+\lambda S^4
+i\bar{\psi}\gamma^{c}V^{\mu}_c(x)[\partial_\mu+\Gamma_\mu(x)]             
\psi -hS\bar{\psi}\psi\right],
\label{1.5z}
\end{eqnarray}                                 
where $h$ and $\lambda$ are dimensionless coupling
constants \cite{footnoteZ3}. As such, the $I_{\rm M}$ action is the most
general curved space matter action for the $\psi(x)$ and $S(x)$ fields
that is invariant under both general coordinate transformations and the local conformal transformation
$S(x)\rightarrow e^{-\alpha(x)}S(x)$, $\psi(x)\rightarrow
e^{-3\alpha(x)/2}\psi(x)$,
$\bar{\psi}(x)\rightarrow e^{-3\alpha(x)/2}\bar{\psi}(x)$,
$V^a_{\mu}(x)\rightarrow e^{\alpha(x)}V^a_{\mu}(x)$,
$g_{\mu\nu}(x)\rightarrow e^{2\alpha(x)}g_{\mu\nu}(x)$. Variation of
this action with respect to 
$\psi(x)$ and
$S(x)$ yields the equations of motion
\begin{eqnarray}
i \gamma^{c}V^{\mu}_c(x)[\partial_{\mu} +\Gamma_\mu(x)]                              
\psi - h S \psi = 0,
\label{1.6z}
\end{eqnarray}                                 
and 
\begin{eqnarray}
\nabla_{\mu}\nabla^{\mu}S+\frac{1}{6}SR^\mu_{\phantom{\mu}\mu}
-4\lambda S^3 +h\bar{\psi}\psi=0.
\label{1.7z}
\end{eqnarray}                                 
We take the fermions (or whatever set of matter fields we may choose, even including the scalar field $S(x)$ itself) to form a general background matter sector perfect fluid, and thus when the scalar field acquires a constant symmetry breaking vacuum expectation value $S_0$ the  total background matter sector energy-momentum tensor is then of the form \cite{Mannheim2006} 
\begin{equation}
T^{\mu \nu} = \frac{1}{c}\left[(\rho_m+p_m)U^{\mu}U^{\nu}+p_mg^{\mu\nu}\right] 
-\frac{1}{6}S_0^2\left(R^{\mu\nu}
-\frac{1}{2}g^{\mu\nu}R^\alpha_{\phantom{\alpha}\alpha}\right)         
-g^{\mu\nu}\lambda S_0^4,
\label{1.8z}
\end{equation}                                 
where the suffix $m$ denotes matter. On taking the background geometry to be the comoving Robertson-Walker metric 
\begin{eqnarray}
ds^2=c^2dt^2-a^2(t)\left[\frac{dr^2}{1-kr^2}+r^2d\theta^2+r^2\sin^2\theta d\phi^2\right]
=c^2dt^2-a^2(t)\tilde{\gamma}_{ij}dx^idx^j,
\label{1.9z}
\end{eqnarray}
where $\tilde{\gamma}^{ij}$ is the background spatial sector metric,
the background $W^{\mu\nu}$ then vanishes, so that the background $T^{\mu\nu}=4\alpha_gW^{\mu\nu}$ then vanishes also. Since the background $T^{\mu\nu}$ does vanish, we can rewrite the $T^{\mu\nu}=0$ equation in the instructive form 
\begin{equation}
\frac{1}{6}S_0^2\left(R^{\mu\nu}
-\frac{1}{2}g^{\mu\nu}R^\alpha_{\phantom{\alpha}\alpha}\right) = 
\frac{1}{c}\left[(\rho_m+p_m)U^{\mu}U^{\nu}+p_mg^{\mu\nu}\right]  -g^{\mu\nu}\lambda S_0^4.
\label{1.10z}
\end{equation}                                 
We thus recognize the conformal cosmological evolution equation given in  (\ref{1.10z}) as being of the form
of none other than the cosmological evolution equation of the standard theory, viz. (on setting $\Lambda =\lambda S_0^4$) 
\begin{eqnarray}
-\frac{c^3}{8\pi G}\left(R^{\mu\nu}
-\frac{1}{2}g^{\mu\nu}R^\alpha_{\phantom{\alpha}\alpha}\right)=\frac{1}{c}\left[(\rho_m+p_m)U^{\mu}U^{\nu}+p_mg^{\mu\nu}\right]  -g^{\mu\nu}\Lambda,
\label{1.11z}
\end{eqnarray}
save only for the fact that the standard $G$ has been replaced by
an effective, dynamically induced one given by
\begin{equation}
G_{{\rm eff}}=-\frac{3c^3}{4\pi S_0^2},
\label{1.12z}
\end{equation}                                 
viz.  by an effective gravitational coupling that,  as noted in \cite{Mannheim1992}, is expressly negative \cite{footnoteZ4}. 
Conformal cosmology is thus 
controlled by an effective gravitational coupling that is
repulsive rather than attractive, and which becomes smaller the larger
$S_0$ might be. With $G_{\rm eff}$ being negative, cosmological gravity is repulsive, and thus naturally leads to cosmic acceleration. 

To see how central the negative sign of $G_{{\rm eff}}$ is to cosmic acceleration 
we define 
\begin{equation}
\bar{\Omega}_{M}(t)=\frac{8\pi G_{{\rm eff}}\rho_{m}(t)}{3c^2H^2(t)}, \quad
\bar{\Omega}_{\Lambda}(t)=\frac{8\pi G_{{\rm
eff}}\Lambda}{3cH^2(t)},\quad \bar{\Omega}_k(t)=-\frac{kc^2}{\dot{a}^2(t)},
\label{1.13z}
\end{equation}                                 
where $H=\dot{a}/a$. And on introducing the deceleration parameter $q=-a\ddot{a}/\dot{a}^2$, from (\ref{1.10z}) we obtain
\begin{eqnarray}
\dot{a}^2(t) +kc^2
&=&\dot{a}^2(t)\left(\bar{\Omega}_{M}(t)+
\bar{\Omega}_{\Lambda}(t)\right),\quad \bar{\Omega}_M(t)+
\bar{\Omega}_{\Lambda}(t)+\bar{\Omega}_k(t)=1,
\nonumber \\
q(t)&=&\frac{1}{2}\left(1+\frac{3p_m}{\rho_m}\right)\bar{\Omega}_M(t)
-\bar{\Omega}_{\Lambda}(t)
\label{1.14z}
\end{eqnarray}
as the background evolution equations of conformal cosmology. 

Without needing to specify any matter sector equation of state and without even needing to solve the theory explicitly at all, we are able to constrain $q(t)$. Specifically, we note that since $\Lambda$ represents the free energy that is released in the phase transition that generated $S_0$ in the first place, $\Lambda $ (and thus the scalar field coupling constant $\lambda$) is necessarily negative. Then with $G_{\rm eff}$ also being negative the quantity $\bar{\Omega}_{\Lambda}(t)$ is positive, i.e., the conformal theory needs a negative $G_{\rm eff}$ in order to obtain a positive $\bar{\Omega}_{\Lambda}(t)$. (In contrast, the standard model rationale for positive $\Omega_{\Lambda}=8\pi G\Lambda/3c^2H^2$ is that since the Newtonian $G$ is positive $\Lambda$ has to be taken to be  positive too.) Since $\rho_m$ and $p_m$ are associated with ordinary matter they are both positive. Thus $\bar{\Omega}_{M}(t)$ is negative and  $\bar{\Omega}_{\Lambda}(t)$ is positive. Thus since $G_{\rm eff}$ is negative it follows that $q(t)$ is automatically negative, being so in every epoch. Consequently, conformal cosmology is automatically accelerating in every cosmological epoch without any adjustment or fine tuning of parameters ever being needed.

While the standard model cannot accommodate a large $\Lambda$ the conformal theory can since $G_{{\rm eff}}$ can be much smaller than $G$. In fact as $S_0$ gets bigger $\Lambda$ gets bigger too but $G_{{\rm eff}}$ gets smaller, with $\bar{\Omega}_{\Lambda}(t)$ self quenching. To see by how much we note that if we set $\Lambda=-a T_V^4/c$ ($V$ denotes vacuum) where $T_V$ is the large temperature at which the $S_0$ generating phase transition occurs,  then with $\bar{\Omega}_M(t)$ being of order $a T^4/c$, in the current era the ratio $\bar{\Omega}_M(t_0)/\bar{\Omega}_{\Lambda}(t_0))=-T_0^4/T_V^4$ is completely negligible. Moreover, since the temperature at recombination is only of order 1 $eV$, at recombination $\bar{\Omega}_M(t)/\bar{\Omega}_{\Lambda}(t)$ is negligible too. Thus we have to go to the very early universe in order to obtain a temperature at which $\bar{\Omega}_M(t)/\bar{\Omega}_{\Lambda}(t)=-T^4/T_V^4$ would not be negligible. In the very early universe we can use $\rho_m=3p_m$ as the equation of state, and while massive matter would be non-relativistic at recombination, it would be irrelevant as to what equation of state we were to use for it since $\bar{\Omega}_M(t)/\bar{\Omega}_{\Lambda}(t)$  is negligible at recombination. Since $\bar{\Omega}_M(t)$ is suppressed  from recombination onward, from recombination until the current era we can replace (\ref{1.14z}) by the simpler and more compact
\begin{eqnarray}
\dot{a}^2(t) +kc^2
&=&\dot{a}^2(t)
\bar{\Omega}_{\Lambda}(t),\quad 
\bar{\Omega}_{\Lambda}(t)+\bar{\Omega}_k(t)=1,\quad
q(t)=-\bar{\Omega}_{\Lambda}(t).
\label{1.15z}
\end{eqnarray}

In order to see by just how much $\bar{\Omega}_{\Lambda}(t)$ might actually be self-quenched numerically in the conformal theory we need to obtain a numerical bound for it. To this end we recall that study of galactic rotation curves in the conformal theory enabled us to determine both the sign and magnitude of $k$, a quantity that is needed for (\ref{1.15z}), and thus for conformal cosmology. Specifically, solving the the fourth-order derivative $W_{\mu\nu}=0$ condition that would hold outside of a static, spherically symmetric star (a situation in which the Weyl tensor is not zero), we find a potential of the form $V^*(r)=-\beta^*c^2/r+\gamma^*c^2r/2$ \cite{Mannheim1989}.  Thus in addition to the Newtonian $1/r$ potential we find a second potential, one  that grows with distance. However, since all the other sources in the Universe would then themselves also be producing  potentials that grow with distance, they would thus also contribute to the net potential that would affect material moving in a given galaxy. As shown in \cite{Mannheim1989,Mannheim2006,Mannheim2011b} this leads to two additional potentials, a linear potential $\gamma_0 c^2r/2$ coming from the cosmological background and a quadratic potential $-\kappa c^2r^2$ coming from the inhomogeneities in that background. With the net potential 
\begin{eqnarray}
V_{\rm TOT}(r)=-\frac{\beta^*c^2}{r}+\frac{\gamma^*c^2r}{2}+\frac{\gamma_0 c^2r}{2}-\kappa c^2r^2,
\label{1.16x}
\end{eqnarray}
conformal gravity is able to fit the rotation curves of 138 galaxies \cite{Mannheim2011b,Mannheim2012c,O'Brien2012}, with  the four parameters $\beta^*$, $\gamma^*$, $\gamma_0$ and $\kappa$ not varying from one galaxy to the next, while being fitted to the universal values
\begin{eqnarray}
\beta^*&=&1.48\times 10^5 {\rm cm},\quad \gamma^*=5.42\times 10^{-41} {\rm cm}^{-1},
\nonumber\\
\gamma_0&=&3.06\times
10^{-30} {\rm cm}^{-1},\quad \kappa = 9.54\times 10^{-54} {\rm cm}^{-2}.
\label{1.17y}
\end{eqnarray} 
In contrast,  we note that for the same 138 galaxy sample dark matter fits need 276 galaxy-dependent free parameters. 

For our purposes here we note that by transforming the comoving Hubble flow to the rest frame of any given galaxy, the universal parameter $\gamma_0$ is found \cite{Mannheim1989} to be related to the spatial three-curvature of the Universe according to $\gamma_0=(-4k)^{1/2}$  \cite{footnoteZ5}. The three-curvature $k$ thus has to be intrinsically negative, with the success of the rotation curve fits showing that the motions of test particles in galaxies measure both the local and global gravitational fields and establish that in conformal cosmology the Universe must be topologically open. Since the net potential $V_{\rm TOT}(r)$ is generated by luminous matter alone, the view of conformal gravity is that the missing mass thought to be needed for rotation curves is not in fact missing at all, it is the rest of the luminous matter in the Universe, and it has just been hiding in plain sight all along.

On taking $k$ to be negative we can now apply the conformal gravity theory to the accelerating universe data. With $k$ being negative it follows that $\bar{\Omega}_k(t)$ is positive. But $\bar{\Omega}_{\Lambda}(t)$ is positive. Thus from (\ref{1.15z}) it follows that $\bar{\Omega}_{\Lambda}(t)$ has to lie between zero and one. Hence no matter how big $\Lambda$ itself might be, $\bar{\Omega}_{\Lambda}(t)$, viz. the amount by which it gravitates, is self-quenched to not be anywhere near as big. Finally from (\ref{1.15z}) it also follows that with no fine tuning at all $q(t)$ is constrained to lie in the narrow negative range $-1\leq q(t)\leq 0$.

Given the above implications of the signs of $G_{\rm eff}$, $\Lambda$ and $k$, we can now solve (\ref{1.15z}) for $a(t)$ so as to determine the evolution of the cosmological background in the region from recombination until the current time $t_0$, and in this region the requisite $a(t)$ is given by \cite{Mannheim2006}
\begin{equation}
a(t)= \frac{(-k)^{1/2}\sinh (\sigma^{1/2} ct)}{\sigma^{1/2}},
\label{1.18y}
\end{equation}
where $\sigma =-2\lambda S_0^2=8\pi G_{\rm eff}\Lambda/3c$ is positive.
With such an $a(t)$ we obtain 
\begin{equation}
\bar{\Omega}_{\Lambda}(t ) = {\rm tanh}^2 (\sigma^{1/2}ct),\quad  \bar{\Omega}_k(t)= {\rm
sech}^2 (\sigma^{1/2}ct), \quad q(t) =-\tanh^2 (\sigma^{1/2}ct),
\label{1.19y}
\end{equation}
and a luminosity distance redshift relation of the form  \cite{Mannheim2006}

\begin{equation} 
d_L=-\frac{c}{H(t_0)}\frac{(1+z)^2}{q_0}\left[1-\left(1+q_0
\frac{q_0}{(1+z)^2}\right)^{1/2}\right],
\label{1.20y}
\end{equation}
where $q_0=q(t_0)$ is the current era value of the deceleration parameter and $H(t_0)$ is the current era value of the Hubble parameter.

Fitting the  type 1A supernovae accelerating universe data with (\ref{1.20y}) gives a fit  \cite{Mannheim2006,Mannheim2012b,Mannheim2017} that is comparable in quality with that of the standard model $\Omega_M=0.3$, $\Omega_{\Lambda}=0.7$ dark matter dark energy paradigm. In the conformal gravity fit $q_0$ is fitted to the value $-0.37$, i.e., quite non-trivially found to be right in the allowed $-1\leq q_0\leq 0$ range, with $\bar{\Omega}_{\Lambda}(t_0)=0.37$, $\bar{\Omega}_{k}(t_0)=0.63$. Since $\bar{\Omega}_M$ is negligible no dark matter is needed, and since $q_0$ and $\bar{\Omega}_{\Lambda}=-q_0$ fall right in the allowed region, no fine tuning is needed either. The ability of the conformal gravity theory to fit the accelerating universe data thus confirms that in conformal cosmology $k$ is indeed negative. And since $k=0$ is favored in the standard concordance cosmological model, it is paramount to ascertain whether conformal cosmological fluctuations can support negative $k$, the ongoing objective of the current conformal cosmological studies \cite{Mannheim2012a,Amarasinghe2019,Phelps2019,Mannheim2020,Amarasinghe2020}.

With ${\rm tanh}^2(\sigma^{1/2}ct_0)$ being fitted to $0.37$, we determine ${\rm tanh}(\sigma^{1/2}ct_0)=0.61$, $\sinh(\sigma^{1/2}ct_0)=0.77$, $\sigma^{1/2}ct_0=0.71$, $H(t_0)=\sigma^{1/2}c/{\rm tanh} (\sigma^{1/2}ct_0)=1.16/t_0$. With $H_0=72$ km/sec/Mpc we obtain $t_0=4.83\times 10^{17}$ sec, a perfectly acceptable value for the age of the universe. Similarly, we obtain $\sigma^{1/2}c=0.15\times 10^{-17}$ ${\rm sec}^{-1}$, $\sigma^{1/2}=0.50\times 10^{-28}$ ${\rm cm}^{-1}$. Recalling that $(-k)^{1/2}=\gamma_0/2=1.53\times 10^{-30}  {\rm cm}^{-1}$, we obtain $a(t_0)=2.36\times10^{-2}$, so the current era expansion radius itself is also small, something that will prove to be central to the study of this paper.

If we extrapolate back to the recombination time $t_R$ we obtain $a(t_R)/a(t_0)=T_0/T_R=O(10^{-4})$. Consequently,  with $\sinh(\sigma^{1/2}ct_0)=0.77$ we obtain  $\sinh(\sigma^{1/2}ct_R)=0.77\times 10^{-4}$. Thus we can approximate $\sinh(\sigma^{1/2}ct_R)$ by $\sigma^{1/2}ct_R$ itself at recombination. Finally then,  to one part in $10^4$ for both $\bar{\Omega}_{\Lambda}(t_R )$ and $ \bar{\Omega}_{k}(t_R )$ we have
\begin{equation} 
a(t_R)=(-k)^{1/2}ct_R,\quad \bar{\Omega}_{\Lambda}(t_R )\approx 0,\quad \bar{\Omega}_{k}(t_R )\approx 1
\label{1.21y}
\end{equation}
at recombination, with $\sigma$ dropping out of $a(t_R)$, and with the numerical value of $a(t_R)$ being $2.36\times 10^{-6}$. As we see, at recombination the conformal universe is curvature dominated. We thus recognize three epochs for conformal cosmology:  radiation dominated early universe, curvature dominated recombination universe, cosmological constant dominated late universe. While there will always be a trace of $\bar{\Omega}_{M}(t)$  in any non-early universe epoch, and while non-early universe propagating matter fields will respond to a geometry that they are not affecting in any substantial way, at recombination we see that $a(t_R)$ as given in (\ref{1.21y}) is  independent not just of $\bar{\Omega}_{M}(t_R)$ but even of $\bar{\Omega}_{\Lambda}(t_R)$ as well. It is the sheer simplicity of (\ref{1.21y}) that  enabled us in \cite{Mannheim2020} to obtain an exact solution to conformal cosmology in the recombination era. In the following we will solve for the cosmology associated with the $a(t)$ given in (\ref{1.18y}), and will again find an exact perturbative solution, one that because the current era $a(t_0)$ is small, actually holds perturbatively all the way from recombination until the current era.

However, before doing so we should note that conformal models in which the scalar field is not an elementary field but actually a vacuum expectation value $\langle \Omega|\bar{\psi}\psi|\Omega\rangle$ of a fermion bilinear have also been considered \cite{Mannheim2012b,Mannheim2017}. In these models it is possible for the matter sources to make a more substantial contribution to cosmic expansion than in the elementary scalar field case \cite{footnoteZ6}.  These dynamical models are not as straightforward to handle as the elementary scalar field model and will be considered elsewhere. And indeed, it is the simplicity of $a(t)= (-k)^{1/2}\sinh (\sigma^{1/2} ct)/\sigma^{1/2}$ in the elementary scalar field model that enables us to solve the model  completely analytically  from recombination onward, just as we now show. Moreover since $a(t)$ is small in the entire region from recombination onward, we can use perturbation theory to determine the gravitational fluctuations in that entire region. Now galaxies will eventually grow to large sizes. However while that will make $\delta \rho_m/\rho_m$ much larger than one, in the elementary scalar field model this overdensity  will still be much smaller than the fluctuations in the gravitational field. Thus once we have solved for the gravitational fluctuations themselves (the objective of this paper), we can then include $\delta \rho_m/\rho_m$ type matter field fluctuations as a small perturbation on them. Moreover, even in the event that matter fields were to contribute substantially to conformal cosmological fluctuations (something our general formalism allows for even though it is not considered here), it is still of value (not just in conformal gravity but even in standard gravity) to ascertain what the pure gravitational contribution itself might be.

\section{The Fluctuations}
\label{S2}

\subsection{Converting the Background to Conformal Time}
\label{S2az}

While the above phenomenological discussion was developed for a specific background conformal cosmology with $k<0$, we now discuss the fluctuation equations for arbitrary $a(t)$, arbitrary $k$ and arbitrary background matter sources. Rather than work in comoving time we have found it more convenient to work in conformal time. Thus on defining
\begin{eqnarray}
\tau =\int \frac{dt}{a(t)},\quad \Omega(\tau)=a(t),
\label{2.1z}
\end{eqnarray}
we replace the background (\ref{1.9z}) by
\begin{eqnarray}
ds^2=\Omega^2(\tau)\left[c^2d\tau^2-\frac{dr^2}{1-kr^2}-r^2d\theta^2-r^2\sin^2\theta d\phi^2\right]
=\Omega^2(\tau)[c^2d\tau^2-\tilde{\gamma}_{ij}dx^idx^j],
\label{2.2z}
\end{eqnarray}
with $\tilde{\gamma}_{ij}$ being the metric of the spatial sector, and with $(i,j,k)=(r,\theta,\phi)$.   In conformal time the background Einstein tensor is given by  
\begin{align}
G_{00}&= -3k- \frac{3}{c^2} \dot{\Omega}^2\Omega^{-2},\quad G_{0i} =0,
\quad G_{ij} = \tilde{\gamma}_{ij}\left[k - \frac{1}{c^2}\dot\Omega^2\Omega^{-2}+ \frac{2}{c^2}\ddot\Omega \Omega^{-1}\right],\quad R^{\alpha}_{\phantom{\alpha}\alpha}=-\frac{6}{\Omega^2}\left[k+ \frac{1}{c^2}\ddot\Omega \Omega^{-1}\right],
\label{2.3z}
\end{align}
where the dot now denotes the derivative with respect to $\tau$. In conformal time  a generic background perfect matter fluid is described by
\begin{align}
T^m_{\mu\nu}&=\frac{1}{c}\left[(\rho_m+p_m)U_{\mu}U_{\nu}+p_mg_{\mu\nu}\right],\quad g^{\mu\nu}U_{\mu}U_{\nu}=-1, \quad U^{0}=\Omega^{-1}(\tau), \quad U_0=-\Omega(\tau), \quad U^{i}=0, \quad U_i=0,
\label{2.4z}
\end{align}
with covariant conservation condition
\begin{align}
 \dot{\rho}_m+3\frac{\dot{\Omega}}{\Omega}(\rho_m+p_m)=0.
\label{2.5z}
\end{align}
For a conformal time radiation fluid with $3p_m=\rho_m$ we obtain $\rho_m=A/\Omega^4$, and for a non-relativistic  fluid with $p_m=0$ we obtain $\rho_m=B/\Omega^3$, viz. the same relations as obtained in comoving time. For conformal cosmology the background evolution equations are of the form
\begin{align}
4\alpha_g W_{\mu\nu}&=\frac{1}{c}\left[(\rho_m+p_m)U_{\mu}U_{\nu}+p_mg_{\mu\nu}\right] 
-\frac{1}{6}S_0^2G_{\mu\nu}         
-g_{\mu\nu}\lambda S_0^4=\Delta^{(0)}_{\mu\nu}, 
\label{2.6z}
\end{align}
with (\ref{2.6z}) serving to define the matter sector $\Delta^{(0)}_{\mu\nu}$.
In a conformal to flat background geometry in which $W_{\mu\nu}=0$ the background evolution equations take the form $\Delta^{(0)}_{\mu\nu}=0$, viz.
\begin{align}
&\frac{1}{2c^2}S_0^2(kc^2 +\dot{\Omega}^2\Omega^{-2})+\frac{\rho_m}{c}\Omega^2+\Omega^2\Lambda=0,\quad
-\frac{1}{6c^2}S_0^2(kc^2 -\dot{\Omega}^2\Omega^{-2}+2\ddot{\Omega}\Omega^{-1})+\frac{p_m}{c}\Omega^2-\Omega^2\Lambda=0,
\nonumber\\
&\frac{1}{3c^2}S_0^2(kc^2 +2\dot{\Omega}^2\Omega^{-2}-\ddot{\Omega}\Omega^{-1})+\frac{\Omega^2}{c}(\rho_m+p_m)=0.
\label{2.7z}
\end{align}
For $\rho_m=A/\Omega^{4}$ we obtain
\begin{align}
-\frac{S_0^2}{2c^2\Lambda}\dot{\Omega}^2=\left[\Omega^2+\frac{kS_0^2}{4\Lambda}+
\left(\frac{k^2S_0^4}{16\Lambda^2}-\frac{A}{\Lambda c}\right)^{1/2}\right]
\left[\Omega^2+\frac{kS_0^2}{4\Lambda}-
\left(\frac{k^2S_0^4}{16\Lambda^2}-\frac{A}{\Lambda c}\right)^{1/2}\right].
\label{2.8z}
\end{align}
While integrating (\ref{2.8z}) gives a somewhat intractable elliptic integral, in the non-early conformal gravity universe we can ignore radiation and set $A=0$, and with $k$ and $\Lambda$ both being negative then obtain 
\begin{align}
\Omega(\tau)=\frac{S_0(k/2\Lambda)^{1/2}}{\sinh(-(-k)^{1/2}c\tau)}=-\frac{(-k/\sigma)^{1/2}}{\sinh((-k)^{1/2}c\tau)}.
\label{2.9z}
\end{align}
To relate the conformal $\tau$ and the comoving $t$, from $a(t)=(-k/\sigma)^{1/2}\sinh(\sigma^{1/2}ct)$ as given in (\ref{1.18y}) we set
\begin{align}
\tau=\int \frac{dt}{(-k/\sigma)^{1/2}\sinh\sigma^{1/2}ct}=\frac{1}{(-kc^2)^{1/2}}\log\tanh(\sigma^{1/2}ct/2),\quad
e^{(-kc^2)^{1/2}\tau}=\tanh(\sigma^{1/2}ct/2),
\label{2.10z}
\end{align}
as normalized so that $\tau=-\infty$ when $t=0$ and $\tau=0$ when $t=\infty$. (With the range of $\tau$ being negative, as given in (\ref{2.9z}) $\Omega(\tau)$ is positive everywhere within the range.) With $\Omega(\tau)=a(t)$,  from (\ref{2.10z}) and $a(t)=(-k/\sigma)^{1/2}\sinh(\sigma^{1/2}ct)$ (\ref{2.9z}) then follows since $\sigma=-2\Lambda/S_0^2$. Finally, since at small comoving $t$ the conformal time $\tau_R$  goes to minus infinity, at recombination we can set $\Omega(\tau_R)=2S_0(k/2\Lambda)^{1/2}\exp[(-k)^{1/2}c\tau_R]$  and $a(t_R)=(-k)^{1/2}ct_R$.

\subsection{The Scalar, Vector, Tensor Basis for Fluctuations}
\label{S2bz}

In analyzing cosmological perturbations it is very convenient to use the scalar, vector, tensor (SVT) basis for the fluctuations as developed in  \cite{Lifshitz1946}  and \cite{Bardeen1980}. In this basis the fluctuations are characterized according to how they transform under three-dimensional rotations, and in this form the basis  has been applied extensively in cosmological perturbation theory (see e.g. \cite{Kodama1984,Mukhanov1992,Stewart1990,Ma1995,Bertschinger1996,Zaldarriaga1998} and \cite{Dodelson2003,Mukhanov2005,Weinberg2008,Lyth2009,Ellis2012}. With the background metric being written with an overall conformal factor $\Omega^2(\tau)$ in (\ref{2.2z}), we shall take the fluctuation metric to also have an overall conformal factor, a particularly convenient choice in the conformal case since both the background $W_{\mu\nu}$ and the perturbative  $\delta W_{\mu\nu}$ transform as $W_{\mu\nu}\rightarrow \Omega^{-2}W_{\mu\nu}$, $\delta W_{\mu\nu}\rightarrow \Omega^{-2}\delta W_{\mu\nu}$  under a conformal transformation, so that, as shown in (\ref{2.20z}), the only dependence of $\delta W_{\mu\nu}$  on $\Omega$ is in an overall  $\Omega^{-2}$ conformal factor. We thus take the full metric to be of the form  \cite{footnoteZ7}
\begin{align}
ds^2&=-(g_{\mu\nu}+h_{\mu\nu})dx^{\mu}dx^{\nu}=\Omega^2(\tau)\left[d\tau^2-\frac{dr^2}{1-kr^2}-r^2d\theta^2-r^2\sin^2\theta d\phi^2\right]
\nonumber\\
&+\Omega^2(\tau)\left[2\phi d\tau^2 -2(\tilde{\nabla}_i B +B_i)d\tau dx^i - [-2\psi\tilde{\gamma}_{ij} +2\tilde{\nabla}_i\tilde{\nabla}_j E + \tilde{\nabla}_i E_j + \tilde{\nabla}_j E_i + 2E_{ij}]dx^i dx^j\right].
\label{2.11z}
\end{align}
In (\ref{2.11z})  $\tilde{\nabla}_i=\partial/\partial x^i$ and  $\tilde{\nabla}^i=\tilde{\gamma}^{ij}\tilde{\nabla}_j$  (with Latin indices) are defined with respect to the background three-space metric $\tilde{\gamma}_{ij}$, and $(1,2,3)=(r,\theta,\phi)$. And with
\begin{eqnarray}
\tilde{\gamma}^{ij}\tilde{\nabla}_j V_i=\tilde{\gamma}^{ij}[\partial_j V_i-\tilde{\Gamma}^{k}_{ij}V_k]
\label{2.12z}
\end{eqnarray}
for any three-vector $V_i$ in a three-space with three-space connection $\tilde{\Gamma}^{k}_{ij}$, the elements of (\ref{2.11z}) are required to obey
\begin{eqnarray}
\tilde{\gamma}^{ij}\tilde{\nabla}_j B_i = 0,\quad \tilde{\gamma}^{ij}\tilde{\nabla}_j E_i = 0, \quad E_{ij}=E_{ji},\quad \tilde{\gamma}^{jk}\tilde{\nabla}_kE_{ij} = 0, \quad\tilde{\gamma}^{ij}E_{ij} = 0.
\label{2.13z}
\end{eqnarray}
With the  three-space sector of the background geometry being maximally three-symmetric, it is described by a Riemann tensor of the form
\begin{eqnarray}
\tilde{R}_{ijk\ell}=k[\tilde{\gamma}_{jk}\tilde{\gamma}_{i\ell}-\tilde{\gamma}_{ik}\tilde{\gamma}_{j\ell}].
\label{2.14z}
\end{eqnarray}
As written, (\ref{2.11z}) contains ten elements, whose transformations are defined with respect to the background spatial sector as four three-dimensional scalars ($\phi$, $B$, $\psi$, $E$) each with one degree of freedom, two transverse three-dimensional vectors ($B_i$, $E_i$) each with two independent degrees of freedom, and one symmetric three-dimensional transverse-traceless tensor ($E_{ij}$) with two degrees of freedom. The great utility of this basis is that since the cosmological fluctuation equations are gauge invariant, only gauge-invariant scalar, vector, or tensor combinations of the components of the scalar, vector, tensor basis can appear in the fluctuation equations. 
In \cite{Amarasinghe2020} it was shown that for the fluctuations associated with the metric given in (\ref{2.11z}) and with $\Omega(\tau)$ being an arbitrary function of $\tau$ and with $k$ also being arbitrary, the gauge-invariant metric combinations are 
\begin{align}
\alpha=\phi + \psi + \dot B - \ddot E,\quad  \gamma= - \dot\Omega^{-1}\Omega \psi + B - \dot E,\quad  B_i-\dot{E}_i,  \quad E_{ij},
\label{2.15z}
\end{align}
for a total of six (one plus one plus two plus two) degrees of freedom, just as required since one can make four coordinate transformations  on the initial ten fluctuation components. As we shall see below, the fluctuation equations will explicitly depend on these specific combinations. Interestingly we note that even with nonzero $k$ the gauge invariant metric combinations have no explicit dependence on $k$.

Given the fluctuation basis we evaluate the fluctuation Einstein tensor, and obtain \cite{Phelps2019} 
\begin{eqnarray}
\delta G_{00}&=& -6 k \phi - 6 k \psi + 6 \dot{\psi} \dot{\Omega} \Omega^{-1} + 2 \dot{\Omega} \Omega^{-1} \tilde{\nabla}_{a}\tilde{\nabla}^{a}B - 2 \dot{\Omega} \Omega^{-1} \tilde{\nabla}_{a}\tilde{\nabla}^{a}\dot{E} - 2 \tilde{\nabla}_{a}\tilde{\nabla}^{a}\psi, 
 \nonumber\\ 
\delta G_{0i}&=& 3 k \tilde{\nabla}_{i}B -  \dot{\Omega}^2 \Omega^{-2} \tilde{\nabla}_{i}B + 2 \overset{..}{\Omega} \Omega^{-1} \tilde{\nabla}_{i}B - 2 k \tilde{\nabla}_{i}\dot{E} - 2 \tilde{\nabla}_{i}\dot{\psi} - 2 \dot{\Omega} \Omega^{-1} \tilde{\nabla}_{i}\phi +2 k B_{i} -  k \dot{E}_{i} \nonumber \\ 
&& -  B_{i} \dot{\Omega}^2 \Omega^{-2} + 2 B_{i} \overset{..}{\Omega} \Omega^{-1} + \frac{1}{2} \tilde{\nabla}_{a}\tilde{\nabla}^{a}B_{i} -  \frac{1}{2} \tilde{\nabla}_{a}\tilde{\nabla}^{a}\dot{E}_{i},
 \nonumber\\ 
\delta G_{ij}&=& -2 \overset{..}{\psi}\tilde{\gamma}_{ij} + 2 \dot{\Omega}^2\tilde{\gamma}_{ij} \phi \Omega^{-2} + 2 \dot{\Omega}^2\tilde{\gamma}_{ij} \psi \Omega^{-2} - 2 \dot{\phi} \dot{\Omega}\tilde{\gamma}_{ij} \Omega^{-1} - 4 \dot{\psi} \dot{\Omega}\tilde{\gamma}_{ij} \Omega^{-1} - 4 \overset{..}{\Omega}\tilde{\gamma}_{ij} \phi \Omega^{-1} \nonumber \\ 
&& - 4 \overset{..}{\Omega}\tilde{\gamma}_{ij} \psi \Omega^{-1} - 2 \dot{\Omega}\tilde{\gamma}_{ij} \Omega^{-1} \tilde{\nabla}_{a}\tilde{\nabla}^{a}B - \tilde{\gamma}_{ij} \tilde{\nabla}_{a}\tilde{\nabla}^{a}\dot{B} +\tilde{\gamma}_{ij} \tilde{\nabla}_{a}\tilde{\nabla}^{a}\overset{..}{E} + 2 \dot{\Omega}\tilde{\gamma}_{ij} \Omega^{-1} \tilde{\nabla}_{a}\tilde{\nabla}^{a}\dot{E} 
\nonumber \\ 
&& - \tilde{\gamma}_{ij} \tilde{\nabla}_{a}\tilde{\nabla}^{a}\phi +\tilde{\gamma}_{ij} \tilde{\nabla}_{a}\tilde{\nabla}^{a}\psi + 2 \dot{\Omega} \Omega^{-1} \tilde{\nabla}_{j}\tilde{\nabla}_{i}B + \tilde{\nabla}_{j}\tilde{\nabla}_{i}\dot{B} -  \tilde{\nabla}_{j}\tilde{\nabla}_{i}\overset{..}{E} - 2 \dot{\Omega} \Omega^{-1} \tilde{\nabla}_{j}\tilde{\nabla}_{i}\dot{E} \nonumber \\ 
&& + 2 k \tilde{\nabla}_{j}\tilde{\nabla}_{i}E - 2 \dot{\Omega}^2 \Omega^{-2} \tilde{\nabla}_{j}\tilde{\nabla}_{i}E + 4 \overset{..}{\Omega} \Omega^{-1} \tilde{\nabla}_{j}\tilde{\nabla}_{i}E + \tilde{\nabla}_{j}\tilde{\nabla}_{i}\phi -  \tilde{\nabla}_{j}\tilde{\nabla}_{i}\psi +\dot{\Omega} \Omega^{-1} \tilde{\nabla}_{i}B_{j} + \frac{1}{2} \tilde{\nabla}_{i}\dot{B}_{j}
\nonumber \\ 
&& -  \frac{1}{2} \tilde{\nabla}_{i}\overset{..}{E}_{j} -  \dot{\Omega} \Omega^{-1} \tilde{\nabla}_{i}\dot{E}_{j} + k \tilde{\nabla}_{i}E_{j} -  \dot{\Omega}^2 \Omega^{-2} \tilde{\nabla}_{i}E_{j} + 2 \overset{..}{\Omega} \Omega^{-1} \tilde{\nabla}_{i}E_{j} + \dot{\Omega} \Omega^{-1} \tilde{\nabla}_{j}B_{i} + \frac{1}{2} \tilde{\nabla}_{j}\dot{B}_{i} \nonumber \\ 
&& -  \frac{1}{2} \tilde{\nabla}_{j}\overset{..}{E}_{i} -  \dot{\Omega} \Omega^{-1} \tilde{\nabla}_{j}\dot{E}_{i} + k \tilde{\nabla}_{j}E_{i} -  \dot{\Omega}^2 \Omega^{-2} \tilde{\nabla}_{j}E_{i} + 2 \overset{..}{\Omega} \Omega^{-1} \tilde{\nabla}_{j}E_{i}- \overset{..}{E}_{ij} - 2 \dot{\Omega}^2 E_{ij} \Omega^{-2} \nonumber \\ 
&& - 2 \dot{E}_{ij} \dot{\Omega} \Omega^{-1} + 4 \overset{..}{\Omega} E_{ij} \Omega^{-1} + \tilde{\nabla}_{a}\tilde{\nabla}^{a}E_{ij},
 \nonumber\\
g^{\mu\nu}\delta G_{\mu\nu} &=& 6 \dot{\Omega}^2 \phi \Omega^{-4} + 6 \dot{\Omega}^2 \psi \Omega^{-4} - 6 \dot{\phi} \dot{\Omega} \Omega^{-3} - 18 \dot{\psi} \dot{\Omega} \Omega^{-3} - 12 \overset{..}{\Omega} \phi \Omega^{-3} - 12 \overset{..}{\Omega} \psi \Omega^{-3} - 6 \overset{..}{\psi} \Omega^{-2} + 6 k \phi \Omega^{-2} \nonumber \\ 
&& + 6 k \psi \Omega^{-2} - 6 \dot{\Omega} \Omega^{-3} \tilde{\nabla}_{a}\tilde{\nabla}^{a}B - 2 \Omega^{-2} \tilde{\nabla}_{a}\tilde{\nabla}^{a}\dot{B} + 2 \Omega^{-2} \tilde{\nabla}_{a}\tilde{\nabla}^{a}\overset{..}{E} + 6 \dot{\Omega} \Omega^{-3} \tilde{\nabla}_{a}\tilde{\nabla}^{a}\dot{E} \nonumber \\ 
&& - 2 \dot{\Omega}^2 \Omega^{-4} \tilde{\nabla}_{a}\tilde{\nabla}^{a}E + 4 \overset{..}{\Omega} \Omega^{-3} \tilde{\nabla}_{a}\tilde{\nabla}^{a}E + 2 k \Omega^{-2} \tilde{\nabla}_{a}\tilde{\nabla}^{a}E - 2 \Omega^{-2} \tilde{\nabla}_{a}\tilde{\nabla}^{a}\phi + 4 \Omega^{-2} \tilde{\nabla}_{a}\tilde{\nabla}^{a}\psi. 
\label{2.16z}
\end{eqnarray}

For fluctuations in the matter field $T^m_{\mu\nu}$ we obtain
\begin{eqnarray}
\delta T^m_{\mu\nu}=\frac{1}{c}\left[(\delta\rho_m+\delta p_m)U_{\mu}U_{\nu}+\delta p_mg_{\mu\nu}+(\rho_m+p_m)(\delta U_{\mu}U_{\nu}+U_{\mu}\delta U_{\nu})+p_mh_{\mu\nu}\right].
\label{2.17z}
\end{eqnarray}
With $g^{\mu\nu}U_{\mu}U_{\nu}=-1$ we obtain 
\begin{eqnarray}
 \delta g^{00}U_{0}U_{0}+2g^{00}U_{0}\delta U_{0}=0,
\label{2.18z}
\end{eqnarray}
which entails that 
\begin{eqnarray}
\delta U_{0}=-\frac{1}{2}(g^{00})^{-1}(-g^{00}g^{00}\delta g_{00})U_{0}=-\Omega(\tau)\phi,
\label{2.19z}
\end{eqnarray}
with $\delta U_0$ thus not being an independent degree of freedom. With $\delta U_i$ being a three-vector we shall decompose it into its transverse and longitudinal parts as $\delta U_i=V_i+\tilde{\nabla}_iV$, where now $\tilde{\gamma}^{ij}\tilde{\nabla}_j V_i=\tilde{\gamma}^{ij}[\partial_j V_i-\tilde{\Gamma}^{k}_{ij}V_k]=0$. As constructed, in general we have 11 fluctuation variables, the six from the metric together with $\delta\rho_m$,  $\delta p_m$ and  the three $\delta U_i$. But we only have ten fluctuation equations. Thus to solve the theory when there is both a $\delta \rho_m$ and a $\delta p_m$ we will need some constraint between $\delta p_m$ and $\delta \rho_m$. However, while this would be required if we want to obtain the general solution, as we had noted above, at recombination and onwards both $\delta p_m$ and $\delta \rho_m$ are suppressed in the conformal case, so no constraint between $\rho_m$ and $p_m$ is needed for our purposes here. Finally, we note that the fluctuation in the cosmological constant term is just  $-\lambda S_0^4 h_{\mu\nu}$.

The fluctuation $\delta W_{\mu\nu}$ in the Bach tensor $W_{\mu\nu}$ is of the form \cite{Amarasinghe2019}
\begin{eqnarray}
\delta W_{00}&=& - \frac{2}{3\Omega^2} (\tilde\nabla_a\tilde\nabla^a + 3k)\tilde\nabla_b\tilde\nabla^b \alpha,
 \nonumber\\ 
\delta W_{0i}&=& -\frac{2}{3\Omega^2}  \tilde\nabla_i (\tilde\nabla_a\tilde\nabla^a + 3k)\dot\alpha
+\frac{1}{2\Omega^2}(\tilde\nabla_b \tilde\nabla^b-\partial_{\tau}^2-2k)(\tilde\nabla_c \tilde\nabla^c+2k)(B_i-\dot{E}_i),
  \nonumber\\ 
\delta W_{ij}&=& -\frac{1}{3 \Omega^2} \left[ \tilde{\gamma}_{ij} \tilde\nabla_a\tilde\nabla^a (\tilde\nabla_b \tilde\nabla^b +2k-\partial_{\tau}^2)\alpha - \tilde\nabla_i\tilde\nabla_j(\tilde\nabla_a\tilde\nabla^a - 3\partial_{\tau}^2)\alpha \right]
\nonumber\\
&& +\frac{1}{2 \Omega^2} \left[ \tilde\nabla_i (\tilde\nabla_a\tilde\nabla^a -2k-\partial_{\tau}^2) (\dot{B}_j-\ddot{E}_j) 
+  \tilde\nabla_j ( \tilde\nabla_a\tilde\nabla^a -2k-\partial_{\tau}^2) (\dot{B}_i-\ddot{E}_i)\right]
\nonumber\\
&&+ \frac{1}{\Omega^2}\left[ (\tilde\nabla_b \tilde\nabla^b-\partial_{\tau}^2-2k)^2+4k\partial_{\tau}^2 \right] E_{ij}.
\label{2.20z}
\end{eqnarray}
We had noted above that since $g^{\mu\nu}\delta W_{\mu\nu}$ does vanish, the tensor $\delta W_{\mu\nu}$ can only have nine independent components. With four coordinate invariances  $\delta W_{\mu\nu}$ can only depend on five (rather than six) gauge invariant degrees of freedom, and as we see, they are $\alpha$, $B_i-\dot{E}_i$ and $E_{ij}$ \cite{footnoteZ8}. As we had also noted above, the only dependence of $\delta W_{\mu\nu}$ on $\Omega$ can be in an overall $\Omega^{-2}$ factor, and thus the five gauge invariant combinations on which it depends must have no explicit dependence on $\Omega$ either (i.e., there must necessarily be five gauge invariant combinations that have no explicit dependence on $\Omega$). And in (\ref{2.15z}) we see that this is explicitly the case, and thus it has to be  the $\Omega$-dependent $\gamma$ that does not appear in $\delta W_{\mu\nu}$ since only the $\Omega$-independent ones can appear. Moreover, even though these same combinations are gauge invariant even if we work in comoving coordinates \cite{Amarasinghe2020}, in comoving coordinates where there is no overall conformal factor in the comoving background metric given in (\ref{1.9z}), the $B_i-\partial_{\tau}E_i$ combination becomes the $a(t)$-dependent $B_i-a(t)\partial_tE_i$. The lack of any explicit dependence of the gauge invariant combinations on $k$ originates in the fact that the Robertson-Walker metric itself can be written in a conformal to flat form (see e.g. \cite{Mannheim2012a}) in which the dependence in $k$ can be put entirely in the conformal factor. Finally, we recall \cite{Amarasinghe2020} that the gauge invariance of the combinations that appear in (\ref{2.15z}) can be established purely kinematically just by looking for combinations that are left invariant under $h_{\mu\nu}\rightarrow h_{\mu\nu}-\nabla_{\mu}\epsilon_{\nu}-\nabla_{\nu}\epsilon_{\mu}$. Consequently, these selfsame combinations must also appear in the fluctuating matter sector energy-momentum tensor $\Delta_{\mu\nu}$ introduced  below (just as they indeed do), and must even be the metric combinations that appear in perturbative Einstein gravity (just as they also in fact do, see e.g. \cite{Amarasinghe2020}).

From (\ref{2.6z}) we obtain  background and fluctuation equations of the form
\begin{align}
4\alpha_g W_{\mu\nu}&=\frac{1}{c}\left[(\rho_m+p_m)U_{\mu}U_{\nu}+p_mg_{\mu\nu}\right] 
-\frac{1}{6}S_0^2G_{\mu\nu}         
-g_{\mu\nu}\Lambda,
\nonumber\\
4\alpha_g \delta W_{\mu\nu}&=\frac{1}{c}\left[(\delta\rho_m+\delta p_m)U_{\mu}U_{\nu}+\delta p_mg_{\mu\nu}+(\rho_m+p_m)(\delta U_{\mu}U_{\nu}+U_{\mu}\delta U_{\nu})+p_mh_{\mu\nu}\right]
-\frac{1}{6}S_0^2\delta G_{\mu\nu} -h_{\mu\nu}\Lambda.
\label{2.21z}
\end{align}
It is convenient to define 
\begin{eqnarray}
\eta=-\frac{24\alpha_g}{S_0^2}, \quad R=-\frac{6(\rho_m+c\Lambda)}{S_0^2}, \quad P=-\frac{6(p_m-c\Lambda)}{S_0^2}, \quad \delta R=-\frac{6\delta \rho_m}{S_0^2}, \quad \delta P=-\frac{6\delta p_m}{S_0^2}. 
\label{2.22y}
\end{eqnarray}
The background and fluctuation equations then take the form
\begin{align}
\eta W_{\mu\nu}&= G_{\mu\nu}+ \frac{1}{c}\left[(R+P)U_{\mu}U_{\nu}+ Pg_{\mu\nu}\right]=\Delta^{(0)}_{\mu\nu},
\label{2.23y}
\end{align}
\begin{align}
\eta \delta W_{\mu\nu}=\delta G_{\mu\nu}+ \frac{1}{c}\left[(\delta R+\delta P)U_{\mu}U_{\nu}+\delta Pg_{\mu\nu}+(R+P)(\delta U_{\mu}U_{\nu}+U_{\mu}\delta U_{\nu})+Ph_{\mu\nu}\right]=\Delta_{\mu\nu},
\label{2.24y}
\end{align}
with (\ref{2.23y}) and (\ref{2.24y}) serving to define $\Delta^{(0)}_{\mu\nu}$ and $\Delta_{\mu\nu}$.
With the use of $\Delta^{(0)}_{\mu\nu}=0$ (which follows here since  $W_{\mu\nu}=0$),  and with $\delta G_{\mu\nu}$ being given in (\ref{2.16z}), the components of $\Delta_{\mu\nu}$ have been determined in \cite{Phelps2019} and are of the form:
\begin{eqnarray}
\Delta_{00}&=& 6 \dot{\Omega}^2 \Omega^{-2}(\alpha-\dot\gamma) + \delta \hat{R} \Omega^2 + 2 \dot{\Omega} \Omega^{-1} \tilde{\nabla}_{a}\tilde{\nabla}^{a}\gamma, 
\label{2.25y}
\end{eqnarray}
\begin{eqnarray}
\Delta_{0i}&=& -2 \dot{\Omega} \Omega^{-1} \tilde{\nabla}_{i}(\alpha - \dot\gamma) + 2 k \tilde{\nabla}_{i}\gamma 
+(-4 \dot{\Omega}^2 \Omega^{-3}  + 2 \overset{..}{\Omega} \Omega^{-2}  - 2 k \Omega^{-1}) \tilde{\nabla}_{i}\hat{V}
\nonumber\\
&& +k(B_i-\dot E_i)+ \frac{1}{2} \tilde{\nabla}_{a}\tilde{\nabla}^{a}(B_{i} - \dot{E}_{i})
+ (-4 \dot{\Omega}^2 \Omega^{-3} + 2 \overset{..}{\Omega} \Omega^{-2} - 2 k \Omega^{-1})V_{i},
\label{2.26y}
\end{eqnarray}
\begin{eqnarray}
\Delta_{ij}&=& \tilde{\gamma}_{ij}\big[ 2 \dot{\Omega}^2 \Omega^{-2}(\alpha-\dot\gamma)
-2  \dot{\Omega} \Omega^{-1}(\dot\alpha -\ddot\gamma)-4\ddot\Omega\Omega^{-1}(\alpha-\dot\gamma)+ \Omega^2 \delta \hat{P}-\tilde\nabla_a\tilde\nabla^a( \alpha + 2\dot\Omega \Omega^{-1}\gamma) \big] 
\nonumber\\
&&+\tilde\nabla_i\tilde\nabla_j( \alpha + 2\dot\Omega \Omega^{-1}\gamma)
+\dot{\Omega} \Omega^{-1} \tilde{\nabla}_{i}(B_{j}-\dot E_j)+\frac{1}{2} \tilde{\nabla}_{i}(\dot{B}_{j}-\ddot{E}_j)
+\dot{\Omega} \Omega^{-1} \tilde{\nabla}_{j}(B_{i}-\dot E_i)+\frac{1}{2} \tilde{\nabla}_{j}(\dot{B}_{i}-\ddot{E}_i)
\nonumber\\
&&- \overset{..}{E}_{ij} - 2 k E_{ij} - 2 \dot{E}_{ij} \dot{\Omega} \Omega^{-1} + \tilde{\nabla}_{a}\tilde{\nabla}^{a}E_{ij},
\label{2.27y}
\end{eqnarray}
\begin{eqnarray}
\tilde{\gamma}^{ij}\Delta_{ij} &=&  6 \dot{\Omega}^2 \Omega^{-2}(\alpha-\dot\gamma)
-6  \dot{\Omega} \Omega^{-1}(\dot\alpha -\ddot\gamma)-12\ddot\Omega\Omega^{-1}(\alpha-\dot\gamma)+ 3\Omega^2 \delta \hat{P}-2\tilde\nabla_a\tilde\nabla^a( \alpha + 2\dot\Omega \Omega^{-1}\gamma),
\label{2.28y}
\end{eqnarray}
\begin{eqnarray}
g^{\mu\nu}\Delta_{\mu\nu}&=& 3 \delta \hat{P} -  \delta \hat{R}
-12 \overset{..}{\Omega}  \Omega^{-3}(\alpha - \dot\gamma) -6 \dot{\Omega} \Omega^{-3}(\dot{\alpha} -\ddot\gamma)
-2 \Omega^{-2} \tilde{\nabla}_{a}\tilde{\nabla}^{a}(\alpha +3\dot\Omega\Omega^{-1}\gamma),
\label{2.29y}
\end{eqnarray}
where
\begin{eqnarray}
\Omega^2 R&=&3k+3\dot{\Omega}^2\Omega^{-2},\quad \Omega^2 P=-k+\dot{\Omega}^2\Omega^{-2}
-2\ddot{\Omega}\Omega^{-1},\quad \dot{R}+3\dot{\Omega}(R+P)\Omega^{-1}=0,
\nonumber\\
\alpha  &=& \phi + \psi + \dot B - \ddot E,\quad \gamma = - \dot\Omega^{-1}\Omega \psi + B - \dot E,\quad \hat{V} = V-\Omega^2 \dot \Omega^{-1}\psi,
 \nonumber\\
\delta \hat{R}&=&\delta R - 12 \dot{\Omega}^2 \psi \Omega^{-4} + 6 \overset{..}{\Omega} \psi \Omega^{-3} - 6 k \psi \Omega^{-2}=\delta R +\dot{\Omega}^{-1}\dot{R}\psi\Omega=\delta R-3(R+P)\psi,
\nonumber\\
\delta \hat{P}&=&\delta P - 4 \dot{\Omega}^2 \psi \Omega^{-4} + 8 \overset{..}{\Omega} \psi \Omega^{-3} + 2 k \psi \Omega^{-2} - 2 \overset{...}{\Omega} \dot{\Omega}^{-1} \psi \Omega^{-2}=\delta P +\dot{\Omega}^{-1}\dot{P}\psi \Omega.
\label{2.30y}
\end{eqnarray}
(The first three expressions in (\ref{2.30y}) hold for the background and follow from $\Delta^{(0)}_{\mu\nu}=0$.) 
With $\eta\delta W_{\mu\nu}-\Delta_{\mu\nu}$ being gauge invariant and with $\delta W_{\mu\nu}$ being gauge invariant on its own, it follows that $\Delta_{\mu\nu}$ is gauge invariant too, and thus its dependence on the metric sector fluctuations must be solely on the metric combinations $\alpha$, $\gamma$, $B_i-\dot{E}_i$ and $E_{ij}$, just as we see. Then since the metric sector $\alpha$, $\gamma$, $B_i-\dot{E}_i$ and $E_{ij}$ are gauge invariant, from the gauge invariance of $\Delta_{\mu\nu}$, and as shown directly in \cite{Amarasinghe2020},   it follows that $\delta \hat{R}$, $\delta \hat{P}$, $\hat{V}$ and $V_i$ are gauge invariant too. We thus have expressed the fluctuation equations entirely in terms of gauge invariant combinations without needing to specify any particular gauge. Given (\ref{2.20z}) and (\ref{2.25y}) - (\ref{2.27y}) the full conformal cosmological fluctuation equations take the form 
\begin{eqnarray}
\eta \delta W_{00}&=& - \frac{2\eta}{3\Omega^2} (\tilde\nabla_a\tilde\nabla^a + 3k)\tilde\nabla_b\tilde\nabla^b \alpha
 \nonumber\\ 
 &=&\Delta_{00}=6 \dot{\Omega}^2 \Omega^{-2}(\alpha-\dot\gamma) + \delta \hat{R} \Omega^2 + 2 \dot{\Omega} \Omega^{-1} \tilde{\nabla}_{a}\tilde{\nabla}^{a}\gamma, 
\label{2.31y}
\end{eqnarray}
\begin{eqnarray}
\eta\delta W_{0i}&=& -\frac{2\eta}{3\Omega^2}  \tilde\nabla_i (\tilde\nabla_a\tilde\nabla^a + 3k)\dot\alpha
+\frac{\eta}{2\Omega^2}(\tilde\nabla_b \tilde\nabla^b-\partial_{\tau}^2-2k)(\tilde\nabla_c \tilde\nabla^c+2k)(B_i-\dot{E}_i)
  \nonumber\\ 
  &=&\Delta_{0i}= -2 \dot{\Omega} \Omega^{-1} \tilde{\nabla}_{i}(\alpha - \dot\gamma) + 2 k \tilde{\nabla}_{i}\gamma 
+(-4 \dot{\Omega}^2 \Omega^{-3}  + 2 \overset{..}{\Omega} \Omega^{-2}  - 2 k \Omega^{-1}) \tilde{\nabla}_{i}\hat{V}
\nonumber\\
&& +k(B_i-\dot E_i)+ \frac{1}{2} \tilde{\nabla}_{a}\tilde{\nabla}^{a}(B_{i} - \dot{E}_{i})
+ (-4 \dot{\Omega}^2 \Omega^{-3} + 2 \overset{..}{\Omega} \Omega^{-2} - 2 k \Omega^{-1})V_{i},
\label{2.32y}
\end{eqnarray}
\begin{eqnarray}
\eta \delta W_{ij}&=& -\frac{\eta}{3 \Omega^2} \left[ \tilde{\gamma}_{ij} \tilde\nabla_a\tilde\nabla^a (\tilde\nabla_b \tilde\nabla^b +2k-\partial_{\tau}^2)\alpha - \tilde\nabla_i\tilde\nabla_j(\tilde\nabla_a\tilde\nabla^a - 3\partial_{\tau}^2)\alpha \right]
\nonumber\\
&& +\frac{\eta}{2 \Omega^2} \left[ \tilde\nabla_i (\tilde\nabla_a\tilde\nabla^a -2k-\partial_{\tau}^2) (\dot{B}_j-\ddot{E}_j) 
+  \tilde\nabla_j ( \tilde\nabla_a\tilde\nabla^a -2k-\partial_{\tau}^2) (\dot{B}_i-\ddot{E}_i)\right]
\nonumber\\
&&+ \frac{\eta}{\Omega^2}\left[ (\tilde\nabla_b \tilde\nabla^b-\partial_{\tau}^2-2k)^2+4k\partial_{\tau}^2 \right] E_{ij}
\nonumber\\
&=&\Delta_{ij}=\tilde{\gamma}_{ij}\big[ 2 \dot{\Omega}^2 \Omega^{-2}(\alpha-\dot\gamma)
-2  \dot{\Omega} \Omega^{-1}(\dot\alpha -\ddot\gamma)-4\ddot\Omega\Omega^{-1}(\alpha-\dot\gamma)+ \Omega^2 \delta \hat{P}-\tilde\nabla_a\tilde\nabla^a( \alpha + 2\dot\Omega \Omega^{-1}\gamma) \big] 
\nonumber\\
&&+\tilde\nabla_i\tilde\nabla_j( \alpha + 2\dot\Omega \Omega^{-1}\gamma)
+\dot{\Omega} \Omega^{-1} \tilde{\nabla}_{i}(B_{j}-\dot E_j)+\frac{1}{2} \tilde{\nabla}_{i}(\dot{B}_{j}-\ddot{E}_j)
+\dot{\Omega} \Omega^{-1} \tilde{\nabla}_{j}(B_{i}-\dot E_i)+\frac{1}{2} \tilde{\nabla}_{j}(\dot{B}_{i}-\ddot{E}_i)
\nonumber\\
&&- \overset{..}{E}_{ij} - 2 k E_{ij} - 2 \dot{E}_{ij} \dot{\Omega} \Omega^{-1} + \tilde{\nabla}_{a}\tilde{\nabla}^{a}E_{ij}.
\label{2.33y}
\end{eqnarray}
In the conformal  gravity theory these cosmological fluctuation equations are completely general, and hold for any possible matter source and any possible $a(t)$ and $k$.

We had noted above that the only difference between the conformal gravity (\ref{1.10z}) and the standard Einstein gravity 
(\ref{1.11z}) was in the replacement of the Newtonian $G$ by the conformal gravity $G_{\rm eff}$ given in (\ref{1.12z}).  We can thus treat both $\Delta^{\mu\nu}_{(0)}$ and $\Delta^{\mu\nu}$ as being generic to both theories. Consequently, Einstein gravity fluctuation theory can be recognized as the $\eta=0$ limit of the conformal gravity $\eta W^{\mu\nu}-\Delta_{(0)}^{\mu\nu}=0$, $\eta \delta W^{\mu\nu}-\Delta^{\mu\nu}=0$  in which $\Delta^{\mu\nu}_{(0)}=0$ and $\Delta^{\mu\nu}=0$.  

We should also add that the parameter $\alpha_g$ is actually known to be negative \cite{Mannheim2011a,Mannheim2016}. The parameter $\eta=-24\alpha_g/S_0^2$ is thus positive. This will prove to be a key feature of the development below as it will lead us to solutions to the fluctuation equations that oscillate in time rather than grow or decay exponentially. 

\section{The Conformal Gravity Decomposition Theorem}
\label{S3}

As constructed, the ten $\eta \delta W_{\mu\nu}=\Delta _{\mu\nu}$ fluctuation equations  mix the perturbative fluctuation quantities. In Einstein gravity a similar situation arises and there one appeals to the decomposition theorem to break the fluctuation equations up into separate scalar, vector and tensor sector equations. By imposing boundary conditions at both $r=\infty$ and $r=0$ a proof of this theorem was given in \cite{Phelps2019} for Einstein gravity. And using the same boundary conditions it was shown in \cite{Mannheim2020} that the decomposition theorem also holds for the conformal gravity fluctuations of interest to us in this paper. Thus for the conformal gravity case the relation $\eta\delta W_{\mu\nu}=\Delta_{\mu\nu}$ breaks up into the ten relations \cite{Mannheim2020} 
\begin{eqnarray}
&& - \frac{2\eta}{3\Omega^2} (\tilde\nabla_a\tilde\nabla^a + 3k)\tilde\nabla_b\tilde\nabla^b \alpha
 =6 \dot{\Omega}^2 \Omega^{-2}(\alpha-\dot\gamma) + \delta \hat{R} \Omega^2 + 2 \dot{\Omega} \Omega^{-1} \tilde{\nabla}_{a}\tilde{\nabla}^{a}\gamma,
\label{3.1x}
 \end{eqnarray}
 \begin{align}
&\frac{1}{2}(\tilde\nabla_c \tilde\nabla^c+2k)\frac{\eta}{\Omega^2}(\tilde\nabla_b \tilde\nabla^b-\partial_{\tau}^2-2k)(B_i-\dot{E}_i)
\nonumber\\
&=\frac{1}{2}(\tilde\nabla_c \tilde\nabla^c+2k)(B_i-\dot{E}_i)
 +(-4 \dot{\Omega}^2 \Omega^{-3} + 2 \overset{..}{\Omega} \Omega^{-2} - 2 k \Omega^{-1})V_{i},
\label{3.2x}
 \end{align}
 \begin{eqnarray}
&& \frac{\eta}{\Omega^2}\left[ (\tilde\nabla_b \tilde\nabla^b-\partial_{\tau}^2-2k)^2+4k\partial_{\tau}^2 \right] E_{ij}
=- \overset{..}{E}_{ij} - 2 k E_{ij} - 2 \dot{E}_{ij} \dot{\Omega} \Omega^{-1} + \tilde{\nabla}_{a}\tilde{\nabla}^{a}E_{ij}.
\label{3.3x}
\end{eqnarray}
 \begin{align}
 &-\frac{\eta}{3 \Omega^2}  \tilde\nabla_a\tilde\nabla^a (\tilde\nabla_b \tilde\nabla^b +2k-\partial_{\tau}^2)\alpha 
 \nonumber\\
 &= 2 \dot{\Omega}^2 \Omega^{-2}(\alpha-\dot\gamma)
-2  \dot{\Omega} \Omega^{-1}(\dot\alpha -\ddot\gamma)-4\ddot\Omega\Omega^{-1}(\alpha-\dot\gamma)+ \Omega^2 \delta \hat{P}-\tilde\nabla_a\tilde\nabla^a( \alpha + 2\dot\Omega \Omega^{-1}\gamma), 
\label{3.4x}
 \end{align}
 \begin{align}
 & \frac{\eta}{3 \Omega^2} (\tilde\nabla_a\tilde\nabla^a - 3\partial_{\tau}^2)\alpha =  \alpha + 2\dot\Omega \Omega^{-1}\gamma,
\label{3.5x}
 \end{align}
 \begin{eqnarray}
&&-\frac{2\eta}{3\Omega^2}  (\tilde\nabla_a\tilde\nabla^a + 3k)\dot\alpha
=-2 \dot{\Omega} \Omega^{-1} (\alpha - \dot\gamma) + 2 k\gamma 
+(-4 \dot{\Omega}^2 \Omega^{-3}  + 2 \overset{..}{\Omega} \Omega^{-2}  - 2 k \Omega^{-1}) \hat{V},
\label{3.6x}
 \end{eqnarray}
\begin{align}
&& \frac{\eta}{2 \Omega^2} (\tilde\nabla_a\tilde\nabla^a -2k-\partial_{\tau}^2) (\dot{B}_i-\ddot{E}_i)
=\dot{\Omega} \Omega^{-1}(B_{i}-\dot E_i)+\frac{1}{2}(\dot{B}_{i}-\ddot{E}_i), 
\label{3.7x}
\end{align}
viz. four one-component scalar equations  [(\ref{3.1x}), (\ref{3.4x}), (\ref{3.5x}), (\ref{3.6x})],  two two-component vector equations [(\ref{3.2x}), (\ref{3.7x})], and one two-component tensor equation [(\ref{3.3x})]. We note that all of these equations are both gauge invariant and exact without approximation, and hold in any cosmological epoch. 

As written, these ten equations involve 11 degrees of freedom, the six associated with the metric, viz. $\alpha$, $\gamma$, $B_i-\dot{E}_i$ and $E_{ij}$, and the five associated with the matter fluctuations, viz. $\delta\hat{R}$, $\delta \hat{P}$, $\hat{V}$ and $V_i$. We note that the tensor equation (\ref{3.3x}) does not involve the fluctuating matter source at all. Thus once we specify a form for $\Omega(\tau)$, something that would involve the background matter source but not the fluctuating one,  we can then solve the tensor sector completely and will do so below. 

Additionally, while the two vector sector equations do involve the matter fluctuation $V_i$, its behavior is highly constrained. Specifically we can write (\ref{3.7x}) as
\begin{align}
\frac{\partial}{\partial\tau}\bigg{[}\eta(\tilde\nabla_a\tilde\nabla^a -2k-\partial_{\tau}^2) (B_i-\dot{E}_i)-\Omega^2(B_i-\dot{E}_i)\bigg{]}=0,
\label{3.8x}
\end{align}
and thus integrate it to 
\begin{align}
\eta(\tilde\nabla_a\tilde\nabla^a -2k-\partial_{\tau}^2) (B_i-\dot{E}_i)-\Omega^2(B_i-\dot{E}_i)=C_i,
\label{3.9x}
\end{align}
where the integration constant $C_i$ depends solely on the spatial coordinates. 
Then, if we apply 
$(\tilde\nabla_c \tilde\nabla^c+2k)$ to (\ref{3.9x}) and compare with (\ref{3.2x}) we obtain a relation that involves $V_i$ alone, viz. 
\begin{eqnarray}
(\tilde\nabla_c \tilde\nabla^c+2k)C_i= 2\Omega^2(-4 \dot{\Omega}^2 \Omega^{-3} + 2 \overset{..}{\Omega} \Omega^{-2} - 2 k \Omega^{-1})V_{i}.
\label{3.10x}
\end{eqnarray}
Then, since $C_i$ is independent of $\tau$, the $\tau$ dependence of $V_i$ is completely fixed, to be  of the unique form $\Omega^{-2}(-4 \dot{\Omega}^2 \Omega^{-3} + 2 \overset{..}{\Omega} \Omega^{-2} - 2 k \Omega^{-1})^{-1}$. Consequently, $V_i$ is not a dynamical variable, and we shall thus set it to zero in the following. With $V_i$ vanishing, $C_i$ then obeys $(\tilde\nabla_c \tilde\nabla^c+2k)C_i=0$. Now in \cite{Phelps2019,Mannheim2020} we showed that the equation $(\tilde\nabla_c \tilde\nabla^c+2k)C_{i}=0$ with any vector $C_i$ had no non-trivial solutions at all that were well behaved at both $r=\infty$ and $r=0$. We thus conclude that $C_i$ is zero, with (\ref{3.9x}) then reducing to
 \begin{align}
\frac{\eta}{\Omega^2}(\tilde\nabla_b \tilde\nabla^b-\partial_{\tau}^2-2k)(B_i-\dot{E}_i)- (B_i-\dot{E}_i)=0.
 \label{3.11y}
 \end{align}
 Alternatively, we could set $C_i=0$ as an initial condition for the integration of (\ref{3.8x}), and then $V_i=0$ would follow from (\ref{3.10x}) unless $-4 \dot{\Omega}^2 \Omega^{-3} + 2 \overset{..}{\Omega} \Omega^{-2} - 2 k \Omega^{-1}$ is zero. And while we will actually take $-4 \dot{\Omega}^2 \Omega^{-3} + 2 \overset{..}{\Omega} \Omega^{-2} - 2 k \Omega^{-1}$ to be zero below (with (\ref{2.9z}) then being its solution), in such a case (\ref{3.10x}) would still lead to  to $C_i=0$. Thus either way we finish up with (\ref{3.11y}).
Thus in the following we only need to solve (\ref{3.11y}), while noting that like the tensor sector (\ref{3.3x}), (\ref{3.11y}) also has no dependence on the matter fluctuations.

Thus the only fluctuation equations that do involve the matter fluctuations are all in the scalar sector.  At this point one cannot in general proceed without further information since while there are four scalar sector equations there are five scalar sector degrees of freedom, $\alpha$, $\gamma$, $\delta \hat{R}$, $\delta \hat{P}$ and $\hat{V}$. To address this issue one looks for some relation between $\delta p_m$ and $\delta \rho_m$. If as is standard we set $\delta p_m/\delta \rho_m=v^2$ and take as equation of state $p_m=w\rho_m$ where $w$ is constant, then from their definitions it follows that 
\begin{eqnarray}
\delta\hat{P}-w\delta \hat{R}=-\frac{6}{S_0^2}(\delta p_m-w\delta \rho_m)=-\frac{6}{S_0^2}(v^2-w)\delta \rho_m.
\label{3.12y}
\end{eqnarray}
Once we specify such $w$ and $v^2$, we can then in principle solve the scalar sector completely in any background cosmology in any cosmological epoch \cite{footnoteZ9}. However, as we had noted above, the specific phenomenological structure of the conformal cosmology of interest to us in this paper is such that the matter sector fluctuation are negligible from recombination onwards, with, as per (\ref{2.7z}),  the expansion radius $\Omega(\tau)$ then obeying $-2 \dot{\Omega}^2 \Omega^{-2}  +  \ddot{\Omega} \Omega^{-1}  - k=0$. Thus with this being case the full set of fluctuation equations is then of the form 
\begin{eqnarray}
&& - \frac{2\eta}{3\Omega^2} (\tilde\nabla_a\tilde\nabla^a + 3k)\tilde\nabla_b\tilde\nabla^b \alpha
 -6 \dot{\Omega}^2 \Omega^{-2}(\alpha-\dot\gamma) -  2 \dot{\Omega} \Omega^{-1} \tilde{\nabla}_{a}\tilde{\nabla}^{a}\gamma=0, 
\label{3.13y}
 \end{eqnarray}
 \begin{align}
 -\frac{\eta}{3 \Omega^2}  \tilde\nabla_a\tilde\nabla^a (\tilde\nabla_b \tilde\nabla^b +2k-\partial_{\tau}^2)\alpha &- 2 \dot{\Omega}^2 \Omega^{-2}(\alpha-\dot\gamma)
+2  \dot{\Omega} \Omega^{-1}(\dot\alpha -\ddot\gamma)
\nonumber\\
&+4\ddot\Omega\Omega^{-1}(\alpha-\dot\gamma)+\tilde\nabla_a\tilde\nabla^a( \alpha + 2\dot\Omega \Omega^{-1}\gamma)=0, 
\label{3.14y}
 \end{align}
 \begin{align}
 & \frac{\eta}{3 \Omega^2} (\tilde\nabla_a\tilde\nabla^a - 3\partial_{\tau}^2)\alpha -  \alpha - 2\dot\Omega \Omega^{-1}\gamma=0,
\label{3.15y}
 \end{align}
 \begin{eqnarray}
&&-\frac{2\eta}{3\Omega^2}  (\tilde\nabla_a\tilde\nabla^a + 3k)\dot\alpha
+2 \dot{\Omega} \Omega^{-1} (\alpha - \dot\gamma) - 2 k\gamma=0,
\label{3.16y}
 \end{eqnarray}
\begin{align}
\frac{\eta}{\Omega^2}(\tilde\nabla_b \tilde\nabla^b-\partial_{\tau}^2-2k)(B_i-\dot{E}_i)-(B_i-\dot{E}_i)=0,
\label{3.17y}
\end{align}
\begin{eqnarray}
&& \frac{\eta}{\Omega^2}\left[ (\tilde\nabla_b \tilde\nabla^b-\partial_{\tau}^2-2k)^2+4k\partial_{\tau}^2 \right] E_{ij}
+ \ddot{E}_{ij} + 2 k E_{ij} + 2 \dot{E}_{ij} \dot{\Omega} \Omega^{-1} - \tilde{\nabla}_{a}\tilde{\nabla}^{a}E_{ij}=0,
\label{3.18y}
\end{eqnarray}
and we note that because $-4 \dot{\Omega}^2 \Omega^{-3}  + 2 \overset{..}{\Omega} \Omega^{-2}  - 2 k \Omega^{-1}$ would now be zero, the $\hat{V}$ term in (\ref{3.6x}) has dropped out identically in (\ref{3.16y}).
These then are the equations that we need to solve.

\section{Scalar Sector Solution from  Recombination until the Current Era}
\label{S4}

For the scalar sector we only have two independent degrees of freedom at recombination, $\alpha$ and $\gamma$, but we have four equations, (\ref{3.13y}), (\ref{3.14y}), (\ref{3.15y}) and (\ref{3.16y}). There must thus be two relations between them, and on setting $-2 \dot{\Omega}^2 \Omega^{-2}  +  \ddot{\Omega} \Omega^{-1}  - k=0$, we find that the left-hand sides of these equations obey 

\begin{align}
&\frac{d}{d\tau}\left(3\dot{\Omega}\Omega^{-1}(\ref{3.16y})+(\ref{3.13y})\right)=\left(\tilde{\nabla}_b\tilde{\nabla}^b+3k-3\dot{\Omega}^2\Omega^{-2}\right)(\ref{3.16y})
-\dot{\Omega}\Omega^{-1}(\ref{3.13y})
+2\dot{\Omega}\Omega^{-1}(\tilde{\nabla}_b\tilde{\nabla}^b+3k)(\ref{3.15y}),
\nonumber\\
&\left(\frac{d}{d\tau} +2\dot{\Omega}\Omega^{-1}\right)(\ref{3.16y})=(\tilde{\nabla}_b\tilde{\nabla}^b+2k)(\ref{3.15y})+(\ref{3.14y})
\label{4.1x}
\end{align}
identically. And with the right-hand sides of (\ref{3.13y}), (\ref{3.14y}), (\ref{3.15y}) and (\ref{3.16y}) vanishing, for the two remaining  relations we note first that
\begin{eqnarray}
&&3\dot{\Omega}\Omega^{-1}(\ref{3.16y})+(\ref{3.13y}) -(\tilde{\nabla}_b\tilde{\nabla}^b+3k)(\ref{3.15y})
=(\tilde{\nabla}_b\tilde{\nabla}^b+3k)\left[\frac{\eta}{\Omega^2}(\ddot{\alpha}-2\dot{\Omega}\Omega^{-1}\dot{\alpha}-\tilde{\nabla}_b\tilde{\nabla}^b\alpha)+\alpha\right]=0,
\label{4.2x}
\end{eqnarray}
and can thus set \cite{footnoteZ10}
\begin{eqnarray}
\eta[\ddot{\alpha}-2\dot{\Omega}\Omega^{-1}\dot{\alpha}-\tilde{\nabla}_b\tilde{\nabla}^b\alpha]=-\Omega^2\alpha,
\label{4.3x}
\end{eqnarray}
to thereby fix $\alpha$. And secondly, from (\ref{3.15y}) we can determine $\gamma$ according to 
 \begin{align}
 \gamma=\frac{\Omega}{2\dot{\Omega}} \left[\frac{\eta}{3 \Omega^2} (\tilde\nabla_a\tilde\nabla^a - 3\partial_{\tau}^2)\alpha -  \alpha\right].
 \label{4.4x}
 \end{align}

As constructed, $\alpha$ will depend on both the spatial coordinates and $\tau$. As discussed in \cite{Phelps2019,Mannheim2020} and in more detail below, the spatial sector separates according to $[\tilde{\nabla}_b\tilde{\nabla}^b+(-k)A_S]\alpha=0$, where $(-k)A_S$ is a separation constant. On introducing  $\rho=(-k)^{1/2}\tau$, and setting $\kappa=\sigma\eta$, then  
with $\Omega(\tau)=-(-k/\sigma)^{1/2}/\sinh((-k)^{1/2}\tau)$ as per (\ref{2.9z}), following the separation of variables the time dependence of $\alpha$, viz. $\alpha(\rho)$, is fixed by
\begin{align}
\left(\frac{d^2}{d\rho^2}+2\frac{\cosh\rho}{\sinh\rho}\frac{d}{d \rho}+A_S+\frac{1}{\kappa \sinh^2\rho}\right)\alpha(\rho)=0,
\label{4.5x}
\end{align}
where $\kappa$ is positive since both $\sigma$ and $\eta$ are positive \cite{footnoteZ11}.

To solve (\ref{4.5x}) we first note that when $\eta$ and $A_S$ are real (below we will see that $A_S$ is real and greater than one), (\ref{4.5x}) is a real second-order derivative equation. Thus its solutions are either real or in complex conjugate pairs. Consequently we can always find real solutions, either ones that are already real or any complex solution plus its complex conjugate. And it is understood that whenever we obtain a complex solution it is always to be accompanied by its complex conjugate. To find the solutions we set $z=\cosh\rho$ so that
\begin{align}
\frac{d}{d\rho}=(z^2-1)^{1/2}\frac{d}{dz},\quad \frac{d^2}{d\rho^2}=(z^2-1)\frac{d^2}{dz^2}+z\frac{d}{dz}.
\label{4.6x}
\end{align}
Then (\ref{4.5x}) becomes
\begin{align}
\left[(z^2-1)\frac{d^2}{dz^2}+3z\frac{d}{dz}+A_S+\frac{1/\kappa}{z^2-1}\right]\alpha=0.
\label{4.7x}
\end{align}
Next we set $\alpha=(z^2-1)^{-1/4}\beta$ 
and obtain
\begin{align}
\left[(z^2-1)\frac{d^2}{dz^2}+2z\frac{d}{dz}+A_S-\frac{3}{4}+\frac{1/\kappa-1/4}{z^2-1}\right]\beta=0.
\label{4.8x}
\end{align}
This is now in the standard form of the associated Legendre function equation
\begin{align}
\left[(z^2-1)\frac{d^2}{dz^2}+2z\frac{d}{dz}-\zeta(\zeta+1)-\frac{\mu^2}{z^2-1}\right]\beta=0,
\label{4.9x}
\end{align}
where, on setting $\nu^2=A_S-1$
\begin{align}
\zeta=-\frac{1}{2}\pm (1-A_S)^{1/2}=-\frac{1}{2}\pm i\nu, \quad \mu^2=\frac{1}{4}-\frac{1}{\kappa},
\label{4.10x}
\end{align}
with associated Legendre solutions $P^{\mu}_{\zeta}(z)$, $Q^{\mu}_{\zeta}(z)$ with $z=\cosh \rho >1$ being given in terms of hypergeometric functions as
\begin{align}
P^{\mu}_{\zeta}(z)=\frac{1}{ \Gamma(1-\mu)}\left(\frac{z+1}{z-1}\right)^{\mu/2}F(-\zeta,\zeta+1;1-\mu;(1-z)/2),
\label{4.11x}
\end{align}
\begin{align}
Q^{\mu}_{\zeta}(z)=\frac{e^{i\mu\pi}\Gamma(\zeta+\mu+1)\Gamma(1/2)}{ 2^{\zeta+1}\Gamma(\zeta+3/2)}(z^2-1)^{\mu/2}z^{-\zeta-\mu-1}
F((\zeta+\mu+2)/2,(\zeta+\mu+1)/2;\zeta+3/2;1/z^2).
\label{4.12x}
\end{align}

\subsection{\bf Conformal Time Scalar Sector Solutions}
\bigskip

Thus in conformal time  the first solution for $\alpha(\rho)$ is of the form 
\begin{align}
\alpha(\rho)=\frac{1}{\sinh^{1/2}\rho}P^{\mu}_{\zeta}(\cosh \rho)= \frac{1}{\sinh^{1/2}\rho}\frac{1}{ \Gamma(1-\mu)}\coth^{\mu}(\rho/2)F(-\zeta,\zeta+1;1-\mu;-\sinh^2(\rho/2)).
\label{4.13x}
\end{align}
We thus obtain the conformal time $\alpha(\rho)$ in a completely closed form.
Once we have $\alpha$ we can get $\gamma$ from (\ref{3.15y}), and with, as per (\ref{2.9z}), $\Omega=-(-k/\sigma)^{1/2}/\sinh\rho$ we have
\begin{align}
-\frac{\kappa}{3}\sinh^2\rho (\nu^2+1+3\partial_{\rho}^2)\alpha(\rho) -  \alpha(\rho) = -2\frac{\cosh\rho}{\sinh\rho}(-k)^{1/2}\gamma(\rho).
\label{4.14x}
\end{align}

\subsection{\bf Comoving Time Scalar Sector Solutions} 
\bigskip

It is also useful to write everything in comoving time. In comoving time $t$ (\ref{4.3x}) becomes
\begin{align}
a^2\frac{d^2\alpha}{dt^2}-a\frac{da}{dt}\frac{d\alpha}{dt}+(-k)A_S\alpha+ \frac{a^2\alpha}{\eta}=0.
\label{4.15x}
\end{align}
Setting $\alpha(t)=\sigma^{3/2}(-k)^{-3/2}a^{3/2}\delta(t)$ we find that $\delta$ obeys
\begin{align}
a^2\frac{d^2\delta}{dt^2}+2a\frac{da}{dt}\frac{d\delta}{dt} +(-k)A_S\delta+ \frac{a^2\delta}{\eta}+\frac{3}{2}a\frac{d^2 a}{dt^2}\delta
-\frac{3}{4}\left(\frac{da}{dt}\right)^2\delta=0.
\label{4.16x}
\end{align}
On setting $\xi=\sigma^{1/2}t$ and, as per (\ref{1.18y}), $a= (-k/\sigma)^{1/2}\sinh \xi$, so that $\alpha=\sinh^{3/2}\xi\delta$,  we obtain
\begin{align}
\ddot{\delta}+2\frac{\cosh \xi}{\sinh \xi}\dot{\delta} +\frac{A_S-3/4}{\sinh^2 \xi}\delta+\left[\frac{1}{\kappa}+\frac{3}{4}\right]\delta=0,
\label{4.17x}
\end{align}
where the dot now denotes $d/d\xi$ and as before $\kappa=\sigma \eta$.
We now proceed as in (\ref{4.5x}) only with the replacements
\begin{align}
A_S\rightarrow  \frac{1}{\kappa}+\frac{3}{4},\quad \frac{1}{\kappa}\rightarrow A_S-\frac{3}{4},
\label{4.18x}
\end{align}
so that now, with $A_S-1=\nu^2$ we obtain
\begin{align}
\zeta=-\frac{1}{2}\pm \left(\frac{1}{4}-\frac{1}{\kappa}\right)^{1/2}, \quad \mu^2=1-A_S=-\nu^2.
\label{4.19x}
\end{align}
Thus we have 
\begin{align}
\alpha(\xi)=\sinh \xi P^{\mu}_{\zeta}(\cosh \xi)
=  \sinh \xi\frac{1}{\Gamma(1-\mu)}\coth^{\mu}(\xi/2)F(-\zeta,\zeta+1;1-\mu;-\sinh^2(\xi/2)).
\label{4.20x}
\end{align}
We thus obtain the comoving time $\alpha(\xi)$ in a completely closed form. 

For $\gamma$, (\ref{4.4x}) is given in comoving time by
\begin{align}
2\frac{d a}{d t} \gamma= \frac{\eta}{3 a^2} \left(\tilde\nabla_a\tilde\nabla^a - 3a^2\frac{\partial^2}{\partial t^2}-3a\frac{da}{dt}\frac{\partial}{\partial t}\right)\alpha -  \alpha,
\label{4.21x}
\end{align}
and thus in terms of $\xi=\sigma^{1/2}t$ and $a=(-k/\sigma)^{1/2}\sinh\xi$ we find that the comoving time dependence of $\gamma$ is given by
\begin{align}
-\frac{\kappa}{3\sinh^{2}\xi}\left(A_S+3\sinh^2\xi\frac{\partial^2}{\partial \xi^2}+3\sinh\xi\cosh\xi\frac{\partial}{\partial \xi}\right)\alpha(\xi) -  \alpha(\xi) = 2\gamma(\xi)(-k)^{1/2}\cosh \xi,
\label{4.22x}
\end{align}
where the comoving time $\alpha(\xi)$ is given in (\ref{4.20x}). 
To evaluate $\gamma(\xi)$ we need to evaluate the first and second time derivatives of $P^{\mu}_{\zeta}(z)$, where $z=\cosh \xi$. To this end we change the variable from $\xi$ to $z$ according to
\begin{align}
 \partial_{\xi}=(z^2-1)^{1/2}\partial_z,\quad \partial_{\xi}^2=(z^2-1)\partial_z^2+z\partial_z.
 \label{4.23x}
\end{align}
Thus with $\sinh \xi=(z^2-1)^{1/2}$ we obtain 
\begin{align}
\sinh^2\xi\partial_{\xi}^2+\sinh\xi\cosh\xi\partial_{\xi}=
(z^2-1)^2\partial_z^2+2z(z^2-1)\partial_z.
\label{4.24x}
\end{align}
Thus from (\ref{4.9x}) we obtain 
\begin{align}
\left[\sinh^2\xi\partial_{\xi}^2+\sinh\xi\cosh\xi\partial_{\xi}\right]P^{\mu}_{\zeta}(z)=\sinh^2\xi\left[\zeta(\zeta+1)+\frac{\mu^2}{\sinh^2\xi}\right]P^{\mu}_{\zeta}(z).
\label{4.25x}
\end{align}
Then with $P^{\mu}_{\zeta}(z)$ obeying the functional relations
\begin{align}
\frac{d}{dz}P^{\mu}_{\zeta}(z)&=(z^2-1)^{-1}\left[(\zeta-\mu+1)P^{\mu}_{\zeta+1}(z)-(\zeta+1)zP^{\mu}_{\zeta}(z)\right],
\nonumber\\
\frac{1}{\sinh\xi}\frac{d}{d \xi}P^{\mu}_{\zeta}(\cosh\xi)&=\frac{1}{\sinh^2\xi}\left[(\zeta-\mu+1)P^{\mu}_{\zeta+1}(\cosh\xi)-(\zeta+1)\cosh\xi P^{\mu}_{\zeta}(\cosh\xi)\right],
\label{4.26x}
\end{align}
for $\alpha(\xi)=\sinh \xi P^{\mu}_{\zeta}(\cosh \xi)$ we obtain 
\begin{align}
&\left[\sinh^2\xi\partial_{\xi}^2+\sinh\xi\cosh\xi\partial_{\xi}\right]\left[\sinh \xi P^{\mu}_{\zeta}(\cosh \xi)\right]=\sinh\xi\Big{[}\zeta(\zeta+1)\sinh^2 \xi +\mu^2 +1+2\sinh^2\xi
\nonumber\\
&-2(\zeta+1)(1+\sinh^2 \xi)\Big{]}P^{\mu}_{\zeta}(\cosh \xi)
+2\sinh \xi\cosh \xi(\zeta-\mu+1)P^{\mu}_{\zeta+1}(\cosh \xi).
\label{4.27x}
\end{align}

Thus finally, for the comoving time  $\gamma(\xi)$ we obtain
\begin{align}
 2\gamma(\xi) (-k)^{1/2}\sinh\xi\cosh \xi &=
-\kappa\Big{[}\zeta(\zeta+1)\sinh^2 \xi +\mu^2 +1+2\sinh^2\xi
-2(\zeta+1)(1+\sinh^2 \xi)\bigg{]}P^{\mu}_{\zeta}(\cosh \xi)
\nonumber\\
&-(\kappa/3) (\nu^2+1)P^{\mu}_{\zeta}(\cosh\xi) -2\kappa\cosh \xi(\zeta-\mu+1)P^{\mu}_{\zeta+1}(\cosh \xi) -\sinh^2\xi P^{\mu}_{\zeta}(\cosh \xi),
\label{4.28x}
\end{align}
where $\mu^2$ and $\zeta$ are given in (\ref{4.19x}).

\section{Vector Sector Solution from  Recombination until the Current Era}
\label{S5}
\bigskip

As we noted above, in the vector sector we need to find solutions to the conformal time (\ref{3.17y}), viz.
\begin{align}
\eta(\partial_{\tau}^2+2k-\tilde{\nabla}_a\tilde{\nabla}^a)(B_i-\partial_{\tau}E_i)
 +\Omega^2(B_i-\partial_{\tau}E_i)=0.
\label{5.1y}
\end{align}
Given that $\Omega =-(-k/\sigma)^{1/2}/\sinh ((-k)^{1/2}\tau)$, on changing the variable to $\rho=(-k)^{1/2}\tau$, on setting  $B_i-\partial_{\tau}E_i=D_i$, and on introducing a separation constant according to $(\tilde{\nabla}_a\tilde{\nabla}^a+(-k)A_V)D_i=0$, we rewrite (\ref{5.1y}) as 
\begin{align}
\frac{\partial^2 D_i}{\partial \tau^2}+(A_V-2)D_i+\frac{D_i}{ \kappa \sinh^2\rho}=0,
\label{5.2y}
\end{align}
where as before $\kappa=\sigma \eta$.
Then on setting $D_i=\sinh\rho C_i$ we obtain 
\begin{align}
\ddot{C_i}+2\frac{\cosh \rho}{\sinh\rho}\dot{C_i}+(A_V-1)C_i+\frac{C_i}{\kappa \sinh^2\rho}=0,
\label{5.3y}
\end{align}
where the dot denotes the derivative with respect to $\rho$.
We recognize this equation as (\ref{4.5x}), and set  $\nu^2=A_V-2$. Thus with
\begin{align}
\zeta=-\frac{1}{2}\pm (2-A_V)^{1/2}=-\frac{1}{2}\pm i\nu, \quad \mu^2=\frac{1}{4}-\frac{1}{\kappa},
\label{5.4y}
\end{align}
the conformal time dependence is given by
\begin{align}
B_i(\rho)-\partial_{\tau}E_i(\rho)=\epsilon_i \sinh^{1/2}\rho P^{\mu}_{\zeta}(\cosh \rho)= \epsilon_i\sinh^{1/2}\rho\frac{1}{\Gamma(1-\mu)}\coth^{\mu}(\rho/2)F(-\zeta,\zeta+1;1-\mu;-\sinh^2(\rho/2)),
\label{5.5y}
\end{align}
where $\epsilon_i$ is a transverse polarization vector.

\subsection{\bf Comoving Time Vector Sector Solutions}

\bigskip

Converting (\ref{5.1y}) to comoving time where now $D_i=B_i-a(t)(\partial E_i/\partial t) $ gives
\begin{align}
\eta((-k)A_V+a^2\partial_t^2+a(\partial_ta)\partial_t+2k)D_i+a^2D_i=0.
 \label{5.6y}
\end{align}
On changing the variable to $\xi=\sigma^{1/2}t$ we obtain 
\begin{align}
\ddot{D_i}+\frac{\dot{a}}{a}\dot{D_i}+\frac{(-k)A_V+2k}{\sigma a^2}D_i+\frac{D_i}{ \sigma\eta}=0,
\label{5.7y}
\end{align}
where the dot denotes the derivative with respect to $\xi$. 
Setting $D_i=\sigma^{1/2}(-k)^{-1/2}a^{1/2}F_i$ gives
\begin{align}
\ddot{F}_i+2\frac{\dot{a}}{a}\dot{F}_i+\frac{(-k)A_V+2k}{\sigma a^2}F_i+\frac{1}{\sigma\eta}F_i
+\frac{\dot{a}^2}{4a^2}F_i+\frac{\ddot{a}}{2a}F_i=0.
\label{5.8y}
\end{align}
Setting $\kappa=\sigma\eta$ and $a=(-k/\sigma)^{1/2}\sinh \xi$, so that $D_i=\sinh^{1/2}\xi F_i$, yields
\begin{align}
\ddot{F}_i+2\frac{\cosh \xi}{\sinh \xi}\dot{F}_i+\frac{A_V-7/4}{\sinh^2\xi}F_i+\left(\frac{1}{ \kappa}+\frac{3}{4}\right)F_i=0.
\label{5.9y}
\end{align}
We now proceed as in (\ref{4.17x}) only with the replacement
\begin{align}
A_S\rightarrow  A_V-1,
\label{5.10y}
\end{align}
so that now with $A_V-2=\nu^2$ we have
\begin{align}
\zeta=-\frac{1}{2}\pm \left(\frac{1}{4}-\frac{1}{\kappa}\right)^{1/2}, \quad \mu^2=2-A_V=-\nu^2.
\label{5.11y}
\end{align}
For the comoving time dependence we thus have 
\begin{align}
B_i(\xi)-a(t)\partial_t E_i(\xi)=\epsilon_iP^{\mu}_{\zeta}(\cosh \xi)
=  \epsilon_i\frac{1}{\Gamma(1-\mu)}\coth^{\mu}(\xi/2)F(-\zeta,\zeta+1;1-\mu;-\sinh^2(\xi/2)),
\label{5.12y}
\end{align}
where $\epsilon_i$ is a transverse polarization vector.

\section{Tensor Sector Solution from Recombination until the Current Era}
\label{S6}

\bigskip

In the tensor sector we have to solve the conformal time (\ref{3.18y}). On setting $\rho=(-k)^{1/2}\tau$ and introducing a tensor sector separation constant that obeys $[\tilde{\nabla}_a\tilde{\nabla}^a+(-k)A_T]E_{ij}=0$, after separating the variables (\ref{3.18y}) takes the form
\begin{align}
-k\eta\Omega^{-2}\left[ (\partial_{\rho}^2-2+A_T)^2-4\partial_{\rho}^2 \right] E_{ij}(\rho)
=-\left[\partial_{\rho}^2-2+2\Omega^{-1}(\partial_{\rho}\Omega)\partial_{\rho}+A_T\right]E_{ij}(\rho).
\label{6.1x}
\end{align}
On setting $\Omega=-(-k/\sigma)^{1/2}/\sinh\rho$ and $\kappa=\sigma\eta$ (\ref{6.1x}) takes the form
\begin{align}
\kappa \sinh^2\rho\left[ (\partial_{\rho}^2-2+A_T)^2-4\partial_{\rho}^2 \right] E_{ij}(\rho)
=-\left[\partial_{\rho}^2-2-2\frac{\cosh\rho}{\sinh\rho}\partial_{\rho}+A_T\right]E_{ij}(\rho).
\label{6.2x}
\end{align}
Symbolically we write (\ref{6.2x})  as $Y=-X$. On setting  $E_{ij}(\rho)=F_{ij}(\rho)\sinh^2\rho$ and dropping the $(i,j)$ indices, on writing $s=\sinh\rho$, $c=\cosh\rho$, we find that following some algebra $X$ and $Y$ are given by
\begin{align}
X=s^2\ddot{F}+2cs\dot{F}-2F-2s^2F+A_Ts^2F,
\label{6.3x}
\end{align}
and
\begin{align}
Y&=\kappa[s^4\overset{....}{F}+8s^3c\overset{...}{F}+(12s^2+16s^4)\ddot{F}-(8s^2+12s^4)F
\nonumber\\
&+2A_Ts^4\ddot{F}+8A_Ts^3c\dot{F}+A_T(4s^2+4s^4)F+A_T^2s^4F]
\nonumber\\
&=\kappa[s^2\ddot{X}+2sc\dot{X}+A_Ts^2X-2s^2X],
\label{6.4x}
\end{align}
where the dot denotes the derivative with respect to $\rho$. We thus find that $Y$ can be expressed as a derivative of $X$, with (\ref{6.2x}) then factorizing into the remarkably compact form
\begin{align}
\left[ \kappa[s^2\partial_{\rho}^2+2sc\partial_{\rho}+A_Ts^2-2s^2]+1\right]X=0.
\label{6.5x}
\end{align}

\subsection{\bf Conformal Time Tensor Sector Solutions}

If we define
\begin{align}
D=\sinh^2\rho\partial_{\rho}^2+2\sinh\rho\cosh\rho\partial_{\rho}+A_T\sinh^2\rho-2\sinh^2\rho-2,
\label{6.6x}
\end{align}
 we can then write (\ref{6.5x}) as
\begin{align}
[\kappa (D+2)+1]DF=0.
\label{6.7x}
\end{align}
To solve (\ref{6.7x}) we identify two classes of solutions:
\begin{align}
DF_1=0,\quad [\kappa (D+2)+1]DF_2=0.
\label{6.8x}
\end{align}
The first solution obeys:
\begin{align}
\left[ \partial_{\rho}^2+2\frac{\cosh\rho}{ \sinh\rho}\partial_{\rho}+A_T-2-\frac{2}{\sinh^2\rho}\right]F_1=0.
\label{6.9x}
\end{align}
Comparing  with (\ref{4.5x})  we replace $A_S$ by $A_T-2$ and $1/\kappa$ by $-2$. Thus on setting $\nu^2=A_T-3$, from the conformal time (\ref{4.10x}) we obtain
\begin{align}
\zeta=-\frac{1}{ 2}\pm (3-A_T)^{1/2}=-\frac{1}{ 2}\pm i\nu, \quad \mu^2=\frac{9}{ 4}.
\label{6.10x}
\end{align}
Consequently, with $\mu=3/2$ $F_1$ is given by
\begin{align}
F_1=(\sinh\rho)^{-1/2} P^{3/2}_{-1/2\pm i\nu}(\cosh\rho)=-\frac{\nu^2\cos\nu\rho}{ \sinh\rho}+\frac{\nu \sin\nu\rho\cosh\rho}{ \sinh^2\rho}.
\label{6.11x}
\end{align}
For $\mu=3/2$ an explicit form exists for $P^{3/2}_{-1/2\pm i\nu}(\cosh\rho)$. It may be found in \cite{Bander1966,Phelps2019,Mannheim2020} and (\ref{7.4z}) below, and we have exhibited it in (\ref{6.11x}).

For the second solution we set $X_2=DF_2$ and rewrite (\ref{6.8x}) as
\begin{align}
 [D+2+1/\kappa]X_2=0,\quad D[X_2+(2+1/\kappa)F_2]=0,
 \label{6.12x}
\end{align}
i.e., 
\begin{align}
\left[\partial_{\rho}^2+2\frac{\cosh\rho}{ \sinh\rho}\partial_{\rho}+A_T-2+\frac{1}{ \kappa\sinh^2\rho}\right]X_2=0,  
\label{6.13x}
\end{align}
\begin{align}
\left[\partial_{\rho}^2+2\frac{\cosh\rho}{\sinh\rho}\partial_{\rho}+A_T-2 -\frac{2}{\sinh^2\rho}\right][X_2+(2+1/\kappa)F_2]=0.  
\label{6.14x}
\end{align}
We recognize (\ref{6.13x}) as (\ref{4.5x}) with $A_T-2=A_S$ and can thus set
\begin{align}
&\zeta=-\frac{1}{2}\pm (3-A_T)^{1/2}=-\frac{1}{2}\pm i\nu,\quad \mu^2=\frac{1}{4}-\frac{1}{\kappa},
\nonumber\\
&X_2=(\sinh\rho)^{-1/2} P^{(1/4-1/\kappa)^{1/2}}_{-1/2\pm i\nu}(\cosh\rho).
\label{6.15x}
\end{align}
On comparing with (\ref{4.5x}), for (\ref{6.14x}) we have $A_S=A_T-2$, $1/\kappa=-2$. Thus we obtain
\begin{align}
&\zeta=-\frac{1}{2}\pm (3-A_T)^{1/2}=-\frac{1}{2}\pm i\nu,\quad \mu^2=\frac{9}{4},
\nonumber\\
&X_2+(2+1/\kappa)F_2=(\sinh\rho)^{-1/2} P^{3/2}_{-1/2\pm i\nu}(\cosh\rho), 
\label{6.16x}
\end{align}
We recognize as $X_2+(2+1/\kappa)F_2$ as $F_1$.
From (\ref{6.16x}) we obtain
\begin{align}
(2+1/\kappa)F_2=(\sinh\rho)^{-1/2} P^{3/2}_{-1/2\pm i\nu}(\cosh\rho) - (\sinh\rho)^{-1/2} P^{(1/4-1/\kappa)^{1/2}}_{-1/2\pm i\nu}(\cosh\rho). 
\label{6.17x}
\end{align}
Given $F_1$ and $F_2$, the two classes of solutions depend on the conformal time as
\begin{align}
E_{ij}(\rho)=A_{ij} \sinh^{3/2}\rho P^{3/2}_{-1/2\pm i\nu}(\cosh\rho) +B_{ij} \sinh^{3/2}\rho P^{(1/4-1/\kappa)^{1/2}}_{-1/2\pm i\nu}(\cosh\rho),
\label{6.18x}
\end{align}
where $A_{ij}$ and $B_{ij}$ are transverse-traceless polarization tensors.

\subsection{\bf Comoving Time Tensor Sector Solutions}
\bigskip

To convert to comoving time we rewrite (\ref{6.9x}), (\ref{6.13x}) and (\ref{6.14x}) as
\begin{align}
\left[\partial_{\rho}^2-2\frac{\partial_{\rho}\Omega}{ \Omega}\partial_{\rho}+A_T-2 -\frac{2\sigma\Omega^2}{(-k)}\right]F_1=0,  
\label{6.19x}
\end{align}
\begin{align}
\left[\partial_{\rho}^2-2\frac{\partial_{\rho}\Omega}{ \Omega}\partial_{\rho}+A_T-2+\frac{\sigma\Omega^2}{(-k)\kappa}\right]X_2=0,
\label{6.20x}
\end{align}
\begin{align}
\left[\partial_{\rho}^2-2\frac{\partial_{\rho}\Omega}{ \Omega}\partial_{\rho}+A_T-2 -\frac{2\sigma\Omega^2}{(-k)}\right][X_2+(2+1/\kappa)F_2]=0. 
\label{6.21x}
\end{align}
Thus in comoving time with $\xi=\sigma^{1/2}t$, $a(t)=(-k/\sigma)^{1/2}\sinh\xi$ and $d/d\rho=\sinh\xi d/d\xi$ we have
\begin{align}
\left[\partial_{\xi}^2-\frac{\cosh \xi }{ \sinh \xi}\partial_{\xi}+\frac{A_T-2}{ \sinh^2\xi} -2\right]F_1=0,  
\label{6.22y}
\end{align}
\begin{align}
\left[\partial_{\xi}^2-\frac{\cosh \xi}{ \sinh \xi}\partial_{\xi}+\frac{A_T-2}{\sinh^2\xi}+\frac{1}{ \kappa}\right]X_2=0,  
\label{6.23y}
\end{align}
\begin{align}
\left[\partial_{\xi}^2-\frac{\cosh {\xi} }{ \sinh \xi}\partial_{\xi}+\frac{A_T-2}{ \sinh^2\xi} -2\right][X_2+(2+1/\kappa)F_2]=0.  
\label{6.24y}
\end{align}

Setting $F_1=\sinh^{3/2}\xi\beta_1$, $X_2=\sinh^{3/2}\xi\beta_2$, $X_2+(2+1/\kappa)F_2=\sinh^{3/2}\xi\beta_3$, $\nu^2=A_T-3$  we obtain
\begin{align}
\left[\partial_{\xi}^2+2\frac{\cosh \xi }{ \sinh \xi}\partial_{\xi}+\frac{\nu^2+1/4}{\sinh^2\xi} -\frac{5}{4}\right]\beta_1=0,  
\label{6.25y}
\end{align}
\begin{align}
\left[\partial_{\xi}^2+2\frac{\cosh \xi}{ \sinh \xi}\partial_{\xi}+\frac{\nu^2+1/4}{ \sinh^2\xi}+\frac{1}{\kappa}+\frac{3}{4}\right]\beta_2=0,  
\label{6.26y}
\end{align}
\begin{align}
\left[\partial_{\xi}^2+2\frac{\cosh \xi }{ \sinh \xi}\partial_{\xi}+\frac{\nu^2+1/4}{ \sinh^2\xi} -\frac{5}{4}\right]\beta_3=0.  
\label{6.27y}
\end{align}
Comparing with (\ref{4.5x}) and (\ref{4.10x}) for $\beta_1$ and $\beta_3$ we obtain
\begin{align}
\frac{1}{\kappa}\rightarrow \nu^2+\frac{1}{4},\quad A_S\rightarrow -\frac{5}{4},\quad \zeta_1=-\frac{1 }{ 2}\pm\frac{3}{2}, \quad \mu^2=-\nu^2,
\label{6.28y}
\end{align}
and for $\beta_2$ we obtain
\begin{align}
\frac{1}{\kappa}\rightarrow \nu^2+\frac{1}{4},\quad A_S\rightarrow \frac{1}{\kappa}+\frac{3}{4},\quad \zeta_2=-\frac{1}{ 2}\pm \left(\frac{1}{4}-\frac{1}{\kappa}\right)^{1/2}, \quad \mu^2=-\nu^2.
\label{6.29y}
\end{align}
Comparing with (\ref{4.13x}), a solution to (\ref{4.5x}) in the coordinates that appear in (\ref{4.5x}), each $\beta$ is given by $\sinh^{-1/2}\xi P^{\mu}_{\zeta}(\cosh\xi)$ with appropriate $\mu$ and $\zeta$. Thus each $F_i$ is given by $\sinh\xi P^{\mu}_{\zeta}(\cosh\xi)$. 
To convert from $F_1$ and $F_2$ to $E_{ij}$ we note that in conformal time we set $E_{ij}=F_{ij}\sinh^2\rho$. With $\Omega(\rho)=(-k/\sigma)^{1/2}/\sinh\rho$ and $a(t)=(-k/\sigma)^{1/2}\sinh\xi$, we set $E_{ij}=F_{ij}(-k/\sigma)/\Omega^2(\rho)=F_{ij}(-k/\sigma)/a^2(\xi)
=F_{ij}/\sinh^2\xi$. Finally then, for $E_{ij}$ the two classes of  solutions depend on the comoving time as 
\begin{align}
E_{ij}(\xi)=\frac{A_{ij}}{ \sinh \xi} P^{\pm i\nu}_{\zeta_1}(\cosh \xi) +\frac{B_{ij}}{ \sinh \xi} P^{\pm i\nu}_{\zeta_2}(\cosh \xi),
\label{6.30y}
\end{align}
where $A_{ij}$ and $B_{ij}$ are transverse-traceless polarization tensors.

Finally, as a check on all of the comoving time solutions, we compare their recombination era behavior with the recombination era behavior that had been determined in \cite{Mannheim2020}. To this end we note with the recombination era being such that $\xi=\sigma^{1/2}t$ is small enough that we can set $\sinh\xi=\xi$, to lowest order in $\xi$ we can set $F(-\zeta,\zeta+1;1-\mu;-\sinh^2(\xi/2))\sim 1+\zeta(\zeta+1)(1-\mu)^{-1}\sinh^2(\xi/2)\sim 1$. Thus from (\ref{4.20x}), (\ref{5.12y}) and (\ref{6.30y}) we obtain $\alpha \sim t^{1\pm i\nu}$, $B_i-a(t)(\partial E_i/\partial t)\sim t^{\pm i\nu}$, $E_{ij}\sim t^{\pm i\nu-1}$  at small $t$, just as had been found at recombination in \cite{Mannheim2020}. We thus establish that with real $\nu$ the solutions oscillate at small time \cite{footnoteZ12}. 

Also we note that for the scalar, vector and tensor solutions we had set $A_S-1=\nu^2$, $A_V-2=\nu^2$, $A_T-3=\nu^2$, with our intent here being that all solutions be associated with one and the same $\nu^2$. By study of the spatial behavior of the solutions we now show that this is in fact the case. Moreover, as the small $t$ solutions already show, for solutions that oscillate in time we need $\nu$ to be real, and thus need the separation constants to obey $A_S\geq 1$, $A_V\geq 2$, $A_T\geq 3$. Similar requirements are met in the spatial sector in order to give spatial oscillations, with the real and continuous parameter $\nu$ being the $k<0$ analog of the familiar real and continuous linear momentum variable $(q_1^2+q_2^2+q_3^2)^{1/2}$ in  a flat three-space.

\section{The Spatial Structure of the Solutions}

To complete the characterization of the fluctuation solutions we need to specify the spatial behavior. Since this had already been discussed in \cite{Phelps2019,Mannheim2020} we briefly state the results. To analyze the spatial behavior of the modes it is convenient to set $r=\sinh\chi/(-k)^{1/2}$, $\tau= \rho/(-k)^{1/2}$, so that the background metric takes the form
\begin{eqnarray}
ds^2=(-k)^{-1}\Omega^2(\tau)\left[d\rho^2-d\chi^2-\sinh^2\chi d\theta^2-\sinh^2\chi\sin^2\theta d\phi^2\right].
\label{7.1z}
\end{eqnarray}
\label{S7}

\subsection{Scalar Sector Spatial Structure}

For the scalar fluctuations described by a generic scalar $S$ we need to solve 
\begin{eqnarray}
\left(\tilde{\nabla}_a\tilde{\nabla}^a+(-k)A_S\right)S=0,
\label{7.2z}
\end{eqnarray}
where $(-k)A_S$ is the separation constant we introduced earlier when discussing the temporal behavior of the solutions.
On setting $S(\chi,\theta,\phi)=S_{\ell}(\chi)Y^m_{\ell}(\theta,\phi)$ (\ref{7.2z}) reduces to 
\begin{eqnarray}
\left[\frac{d^2}{d\chi^2}+2\frac{\cosh\chi }{\sinh\chi}\frac{d }{ d\chi}
-\frac{\ell(\ell+1)}{ \sinh^2\chi}+A_S\right]S_{\ell}(\chi)=0.
\label{7.3z}
\end{eqnarray}
On making the identification $\ell(\ell+1)=-1/\kappa$, we recognize this equation as being none other than the same fluctuation equation (\ref{4.5x}) that had met earlier in studying the time behavior. Now while we could identify the solution as $P^{\mu}_{\zeta}(\cosh\chi)$ with $\zeta=-1/2\pm i\nu$, $\mu^2=(\ell+1/2)^2$, because $\ell$ is an integer (\ref{7.3z}) can be solved directly in terms of elementary functions. Specifically, there are two classes of solutions to this second-order differential equation, labelled $\hat{S}_{\ell}$ and $\hat{f}_{\ell}\hat{S}_{\ell}$, and they are of the form 
\begin{eqnarray}
\hat{S}_{\ell}=\sinh^{\ell}\chi\left(\frac{1}{ \sinh\chi} \frac{d }{ d\chi}\right)^{\ell+1}\cos\nu\chi,\quad \hat{f}_{\ell}\hat{S}_{\ell}=\hat{S}_{\ell}\int \frac{d\chi }{\sinh^2\chi\hat{S}_{\ell}^2},\quad \nu^2=A_S-1,
\label{7.4z}
\end{eqnarray}
with both solutions being even in $\nu$.
The $\hat{S}_{\ell}$ solution is given in \cite{Bander1966,Phelps2019,Mannheim2020} and the $\hat{f}_{\ell}\hat{S}_{\ell}$ solution is given in \cite{Phelps2019,Mannheim2020}. Both of these sets of solutions with real and continuous  $\nu^2$ (i.e., real and continuous $A_S>1$) are complete. With the asymptotic behavior of the solutions being of the form $e^{-\chi}e^{\pm i\nu\chi}$ \cite{Phelps2019,Mannheim2020}, the solutions indeed oscillate if $\nu$ is real. Moreover, at $\chi=0$ the solutions behave as $\chi^{\ell}$ or as $\chi^{-\ell-1}$, with there always being one solution that is bounded as $e^{-\chi}e^{\pm i\nu\chi}$  at $\chi=\infty$ and well behaved at $\chi=0$ for all $\ell$, and one solution that is bounded as $e^{-\chi}e^{\pm i\nu\chi}$ at $\chi=\infty$ but badly behaved at $\chi=0$ for all $\ell$.  Typical examples are $\hat{S}_{\ell=0}=-\nu\sin\nu\chi/\sinh\chi$, $\hat{f}_{\ell=0}\hat{S}_{\ell=0}=\cos\nu\chi/\nu^2\sinh\chi$. Both solutions are bounded at $\chi=\infty$, with the first solution behaving as $\chi^0$ at $\chi=0$ and the other as $\chi^{-1}$.

In  \cite{Bander1966} a master equation was introduced that provides normalization factors for scalar sector modes that are propagating in a space with $F$ spatial dimensions. We introduce 
\begin{eqnarray}
Z_{\nu,\beta}(\chi)=A(\nu,\beta,F)\sinh^{\beta}\chi\left(\frac{1}{\sinh\chi}\frac{d}{d\chi}\right)^{(F-1)/2+\beta}\cos\nu\chi,
\label{7.5y}
\end{eqnarray}
with $\ell=\beta +(F-3)/2$ and 
\begin{eqnarray}
A(\nu,\beta,F)=\frac{2^{1/2}}{[\pi \nu^2(\nu^2+1^2)(\nu^2+2^2)......(\nu^2+((F-3)/2+\beta)^2)]^{1/2}}.
\label{7.6y}
\end{eqnarray}
With the $\chi$ dependence of $F$-dimensional integration measure being $\sinh^{F-1}\chi$, the modes obey the Dirac delta function orthonormality condition \cite{Bander1966}
\begin{eqnarray}
\int_0^{\infty}d\chi\sinh^2\chi \sinh^{(F-3)/2}\chi Z_{\nu_1,\beta_1}(\chi) \sinh^{(F-3)/2}\chi Z^*_{\nu_2,\beta_2}(\chi)=\delta_{\beta_1,\beta_2}\delta(\nu_1-\nu_2).
\label{7.7y}
\end{eqnarray}
Thus for the dimension $F=3$ that we are interested in we have
\begin{eqnarray}
\int_0^{\infty}d\chi\sinh^2\chi Z_{\nu_1,\ell_1}(\chi) Z^*_{\nu_2,\ell_2}(\chi)=\delta_{\ell_1,\ell_2}\delta(\nu_1-\nu_2).
\label{7.8y}
\end{eqnarray}
In regards to this normalization condition we note that since maximally  symmetric three-spaces with metric $d\chi^2+\sinh^2\chi d\theta^2+\sinh^2\chi \sin^2\theta d\phi^2$ and integration measure $\sinh^2\chi\sin\theta$ are conformal to flat, the delta function normalization that is needed for plane waves propagating in a flat space translates into the delta function normalization for the $Z_{\nu,\beta}(\chi)$ that is given here, with the continuous parameter $\nu$ replacing the familiar continuous linear momentum variable.

\subsection{Vector Sector Spatial Structure}

For the vector fluctuations described by a generic vector $V_i$ we need to solve 
\begin{eqnarray}
\left(\tilde{\nabla}_a\tilde{\nabla}^a+(-k)A_V\right)V_i=0,
\label{7.9y}
\end{eqnarray}
where $(-k)A_V$ is the separation constant we introduced earlier. While the various components of $V_i$ are mixed in (\ref{7.9y}), as shown in \cite{Phelps2019}, for a transverse $V_i$ that obeys $\tilde{\nabla}_iV^i=0$ the equation for $V_1$ only involves $V_1$ (i.e., $V_{\chi}$). And on 
 setting $V_1(\chi,\theta,\phi)=\hat{V}_{\ell}(\chi)Y^m_{\ell}(\theta,\phi)$ the equation for $V_1$  reduces to 
\begin{eqnarray}
\left[\frac{d^2}{d\chi^2}+4\frac{\cosh\chi}{ \sinh\chi}\frac{d }{d\chi}
+2+A_V+\frac{2 }{ \sinh^2\chi}-\frac{\ell(\ell+1)}{ \sinh^2\chi}\right]\hat{V}_{\ell}(\chi)=0.
\label{7.10y}
\end{eqnarray}
On setting $\hat{V}_{\ell}=\alpha_{\ell}/\sinh\chi$ we  find that (\ref{7.10y}) takes the form
\begin{eqnarray}
\frac{1}{L^2}\left[\frac{d^2 }{d\chi^2}+2\frac{\cosh\chi}{ \sinh\chi}\frac{d }{d\chi}
-\frac{\ell(\ell+1) }{\sinh^2\chi}+A_V-1\right]\alpha_{\ell}=0.
\label{7.11y}
\end{eqnarray}
We recognize (\ref{7.11y}) as being in the same form as  (\ref{7.3z}), but with $A_S$ replaced by $A_V-1$ so that $\nu^2=A_V-2$.

Solutions to (\ref{7.10y}) behave asymptotically as $e^{-2\chi}e^{\pm i\nu\chi}$, and at $\chi=0$ as $\chi^{\ell-1}$, $\chi^{-\ell-2}$. Thus the solutions indeed oscillate if $\nu^2$ is real, and with real and continuous $\nu^2$ (i.e., real and continuous $A_V>2$) the solutions are complete. Moreover, at $\chi=0$ there is always one solution that is bounded as $e^{-2\chi}e^{\pm i\nu\chi}$  at $\chi=\infty$ and well behaved at $\chi=0$ for all $\ell \geq 1$, and one solution that is bounded as $e^{-2\chi}e^{\pm i\nu\chi}$ at $\chi=\infty$ but badly behaved at $\chi=0$ for all $\ell \geq 1$. A typical example of a solution that behaves as $e^{-2\chi}e^{\pm i\nu\chi}$  asymptotically and is well-behaved at $\chi=0$  is  $\hat{V}_{\ell=1}=\nu\sin\nu\chi\cosh\chi/\sinh^3\chi-\nu^2\cos\nu\chi/\sinh^2\chi$.

For the normalization of the vector modes we recall that while the normalization condition given in (\ref{7.7y}) was established for scalar modes in spaces with dimension $F$, it was noted in \cite{Mannheim2020} that these same conditions hold for vector modes in three dimensions if we set $F=5$, $\ell=\beta+1$ in (\ref{7.7y}). Thus for vector modes in three  spatial dimensions  we obtain
\begin{eqnarray}
\int_0^{\infty}d\chi\sinh^2\chi \sinh\chi Z_{\nu_1,\ell_1-1}(\chi) \sinh\chi Z^*_{\nu_2,\ell_2-1}(\chi)=\delta_{\ell_1,\ell_2}\delta(\nu_1-\nu_2).
\label{7.12y}
\end{eqnarray}
As we see, compared with the scalar (\ref{7.8y}) the vector sector  measure includes two extra factors of $\sinh\chi$ in order to balance the replacement of $\hat{V}_{\ell}$ by $\alpha_{\ell}/\sinh\chi$, so that asymptotically for a scalar that behaves as $e^{-\chi}e^{\pm i\nu}$ the vector behaves as $e^{-2\chi}e^{\pm i\nu}$. Moreover, the change in $\beta$ by one unit shifts $\ell$ by one unit so that only $\ell \geq 1$ is encompassed in (\ref{7.12y}), just as is required to have the vector modes be well behaved at $\chi=0$.

Using these same techniques and the formalism presented in \cite{Phelps2019,Mannheim2020} we could solve for $V_2$ and $V_3$ as well, but do not actually need to do so explicitly since for study of the anisotropy in the cosmic microwave background only line of sight radial mode fluctuations are detectable by a current era observer (see e.g. \cite{Weinberg2008}).

\subsection{Tensor Sector Spatial Structure}

For the tensor fluctuations described by a generic tensor $T_{ij}$ we need to solve 
\begin{eqnarray}
\left(\tilde{\nabla}_a\tilde{\nabla}^a+(-k)A_T\right)T_{ij}=0,
\label{7.13y}
\end{eqnarray}
where $(-k)A_T$ is the separation constant we introduced earlier. While the various components of $T_{ij}$ are mixed in (\ref{7.13y}), as shown in \cite{Phelps2019} for a transverse-traceless $T_{ij}$ the equation for $T_{11}$ only involves $T_{11}$ (viz. $T_{\chi,\chi}$). And on 
 setting $T_{11}(\chi,\theta,\phi)=\hat{T}_{\ell}(\chi)Y^m_{\ell}(\theta,\phi)$ the equation for $T_{11}$  reduces to 
\begin{eqnarray}
\left[\frac{d^2}{d\chi^2}+6\frac{\cosh\chi}{ \sinh\chi}\frac{d }{d\chi}
+6+\frac{6 }{ \sinh^2\chi}-\frac{\ell(\ell+1)}{ \sinh^2\chi}+A_T\right]\hat{T}_{\ell}(\chi)=0.
\label{7.14y}
\end{eqnarray}
On setting $\hat{T}_{\ell}=\gamma_{\ell}/\sinh^2\chi$ we  find that (\ref{7.14y}) takes the form
\begin{eqnarray}
\left[\frac{d^2}{d\chi^2}+2\frac{\cosh\chi}{\sinh\chi}\frac{d}{d\chi}
-\frac{\ell(\ell+1) }{ \sinh^2\chi}-2+A_T\right]\gamma_{\ell}=0.
\label{7.15y}
\end{eqnarray}
We recognize (\ref{7.15y}) as being in the same form as  (\ref{7.3z}), but with $A_S$ replaced by $A_T-2$ so that $\nu^2=A_T-3$.

Solutions to (\ref{7.14y}) behave asymptotically as $e^{-3\chi}e^{\pm i\nu\chi}$, and at $\chi=0$ as $\chi^{\ell-2}$, $\chi^{-\ell-3}$. Thus the solutions indeed oscillate if $\nu$ is real, and with real and continuous $\nu^2$ (i.e., real and continuous $A_T>3$) the solutions are complete. Moreover, at $\chi=0$ there will always be one solution that is bounded as $e^{-3\chi}e^{\pm i\nu\chi}$  at $\chi=\infty$ and well behaved at $\chi=0$ for all $\ell \geq 2$, and one solution that is bounded as $e^{-3\chi}e^{\pm i\nu\chi}$ at $\chi=\infty$ but badly behaved at $\chi=0$ for all $\ell \geq 2$. A typical example of a solution that behaves as $e^{-3\chi}e^{\pm i\nu\chi}$ asymptotically and is well-behaved at $\chi=0$  is  $\hat{T}_{\ell=2}=3\nu^2\cos\nu\chi\cosh\chi/\sinh^4\chi-\nu(2-\nu^2)\sin\nu\chi/\sinh^3\chi-3\nu\sin\nu\chi/\sinh^5\chi$.

For the normalization of the tensor modes we recall that while the normalization condition given in (\ref{7.7y}) was established for scalar modes in spaces with dimension $F$, it was noted in \cite{Mannheim2020} that these same conditions hold for tensor modes in three dimensions if we set $F=7$, $\ell=\beta+2$ in (\ref{7.7y}). Thus for tensor modes in three spatial dimensions we obtain
\begin{eqnarray}
\int_0^{\infty}d\chi\sinh^2\chi \sinh^2\chi Z_{\nu_1,\ell_1-2}(\chi) \sinh^2\chi Z^*_{\nu_2,\ell_2-2}(\chi)=\delta_{\ell_1,\ell_2}\delta(\nu_1-\nu_2).
\label{7.16y}
\end{eqnarray}
As we see, compared with the scalar (\ref{7.8y}) the tensor sector  measure includes two extra factors of $\sinh^2\chi$ in order to balance the replacement of $T_{11,\ell}$ by $\gamma_{\ell}/\sinh^2\chi$, so that asymptotically for a scalar that behaves as $e^{-\chi}e^{\pm i\nu}$ the tensor behaves as $e^{-3\chi}e^{\pm i\nu}$. Moreover, the change in $\beta$ by two units shifts $\ell$ by two units so that only $\ell \geq 2$ are encompassed in (\ref{7.16y}), just as is required to have the tensor modes be well behaved at $\chi=0$.

Using these same techniques and the formalism presented in \cite{Phelps2019,Mannheim2020} we could solve for the other components of $T_{ij}$ as well, but as with the vector fluctuations, we do not actually need to do so explicitly since only line of sight radial mode fluctuations in the cosmic microwave background are detectable by a current era observer.

\section{The Full Solution from Recombination to the Current Era}
\label{S8}

Putting everything together, in conformal cosmology the full comoving time radial mode solutions from recombination to the current era are of the form 
\begin{align}
\alpha&= A(\nu,\ell,3)\hat{S}_{\ell}(\chi)Y^m_{\ell}(\theta,\phi)\sinh (\sigma^{1/2}t)P^{\mu}_{\zeta}(\cosh(\sigma^{1/2}t)),\quad
\zeta=-\frac{1}{2}\pm \left(\frac{1}{4}-\frac{1}{\sigma\eta}\right)^{1/2}, \quad \mu^2=1-A_S=-\nu^2,
\label{8.1z}
\end{align}
\begin{align}
2\frac{d a}{d t} \gamma= \frac{\eta}{3 a^2} \left(\tilde\nabla_a\tilde\nabla^a - 3a^2\frac{\partial^2}{\partial t^2}-3a\frac{da}{dt}\frac{\partial}{\partial t}\right)\alpha -  \alpha,
\label{8.2z}
\end{align}
\begin{align}
\nonumber\\
B_1-a(t)\partial_t E_1&=\epsilon_1A(\nu,\ell -1,5)\hat{V}_{\ell}(\chi)Y^m_{\ell}(\theta,\phi)P^{\mu}_{\zeta}(\cosh(\sigma^{1/2}t)),\quad
\zeta=-\frac{1}{2}\pm \left(\frac{1}{4}-\frac{1}{\sigma\eta}\right)^{1/2}, \quad \mu^2=2-A_V=-\nu^2,
\label{8.3z}
\end{align}
\begin{align}
E_{11}&=A(\nu,\ell -2,7)\hat{T}_{\ell}(\chi)Y^m_{\ell}(\theta,\phi)\frac{1}{\sinh(\sigma^{1/2}t)}\bigg{[}
A_{11}P^{\mu}_{\zeta_1}(\cosh(\sigma^{1/2}t)+B_{11}P^{\mu}_{\zeta_2}(\cosh(\sigma^{1/2}t)\bigg{]},
\nonumber\\
\zeta_1&=-\frac{1}{2}\pm \frac{3}{2}, \quad \zeta_2=-\frac{1}{2}\pm \left(\frac{1}{4}-\frac{1}{\sigma\eta}\right)^{1/2},\quad \mu^2=3-A_T=-\nu^2,
\label{8.4z}
\end{align}
for all real and positive $\nu^2$. In the conformal theory these radial scalar, vector and tensor mode solutions are exact without approximation from recombination until the current era.

\begin{acknowledgments}
One of us (PDM)  acknowledges a useful conversation with Dr. J. Erlich.
\end{acknowledgments}


\begin{thebibliography}{99}

\bibitem{Dodelson2003} S. Dodelson, \textit{Modern Cosmology} (Academic Press, 2003).

\bibitem{Mukhanov2005} V. Mukhanov, \textit{Physical Foundations of Cosmology} (Cambridge University Press, Cambridge U. K. 2005).

\bibitem{Weinberg2008} S. Weinberg, \textit{Cosmology} (Oxford University Press, Oxford U. K. 2008).

\bibitem{Lyth2009} D. H. Lyth and A. R. Liddle, \textit{The Primordial Density Perturbation: Cosmology, Inflation and the Origin of Structure} (Cambridge University Press, Cambridge U. K. 2009).

\bibitem{Ellis2012} G. F. R. Ellis, R. Maartens and  M. A. H. MacCallum, {\it Relativistic Cosmology} (Cambridge University Press,  Cambridge U. K. 2012).

\bibitem{Guth1981}  \href{https://doi.org/10.1103/PhysRevD.23.347} {A. H. Guth, Phys. Rev. D \textbf{23}, 347 (1981).}


\bibitem{Bahcall2000}\href{https://doi.org/10.1126/science.284.5419.1481} {N. A. Bahcall, J. P. Ostriker, S. Perlmutter, 
and P. J. Steinhardt, Science \textbf {284}, 1481 (1999).}

\bibitem{deBernardis2000}\href{https://doi.org/10.1038/35010035}  {P. de Bernardis et. al., Nature \textbf{404}, 955 (2000).}

\bibitem{Tegmark2004} \href{https://doi.org/10.1103/PhysRevD.69.103501} {M. Tegmark et. al., Phys. Rev. D \textbf{69}, 103501(2004).}

\bibitem{footnoteZ1} That quantum mechanics is even relevant on large macroscopic scales is already manifest in the cosmic microwave background itself as it is described by an intrinsically quantum-mechanical  black-body radiation spectrum with an $\hbar$-dependent energy density $\pi^2k_{\rm B}^4T^4/15\hbar^3c^3$. Also, the stability of both white dwarf stars and neutron stars is provided by the Fermi-Dirac statistics of the spin one-half particles in the star, with the Chandrasekhar mass $(\hbar c/G)^{3/2}/m_p^2$ of white dwarf stars intrinsically depending on $\hbar$.

\bibitem{Mannheim1990} \href{https://doi.org/10.1007/BF00756278}{P. D. Mannheim, Gen. Rel. Gravit. {\bf 22}, 289 (1990).}

\bibitem{Mannheim1989} \href{https://doi.org/10.1086/167623}{P. D. Mannheim and D. Kazanas, Astrophys. J. \textbf{342}, 635 (1989).} 

\bibitem{Mannheim1994} \href{https://doi.org/10.1007/BF02105226}{P. D. Mannheim and D. Kazanas, Gen. Rel. Gravit. \textbf{26}, 337 (1994).}


\bibitem{Mannheim2006} \href{https://doi.org/10.1016/j.ppnp.2005.08.001}{P. D. Mannheim, Prog. Part. Nucl. Phys. \textbf{ 56}, 340 (2006). }

\bibitem{Mannheim2012b} \href{https://doi.org/10.1007/s10701-011-9608-6}{P. D. Mannheim, Found. Phys. \textbf{42}, 388 (2012).}

\bibitem{Mannheim2017}  \href{https://doi.org/10.1016/j.ppnp.2017.02.001}{P. D. Mannheim, Prog. Part. Nucl. Phys. \textbf{94}, 125 (2017).}

\bibitem{Mannheim2011b} \href{https://doi.org/10.1103/PhysRevLett.106.121101}{P. D. Mannheim and J. G. O'Brien,  Phys. Rev. Lett. \textbf{106}, 121101 (2011).}

\bibitem{Mannheim2012c} \href{https://doi.org/10.1103/PhysRevD.85.124020}{P. D. Mannheim and J. G. O'Brien, Phys. Rev. D \textbf{85}, 124020 (2012).}

\bibitem{O'Brien2012} \href{https://doi.org/10.1111/j.1365-2966.2011.20386.x}{J. G. O'Brien and P. D. Mannheim, Mon. Not. R. Astron. Soc. \textbf{421,} 1273 (2012).}

\bibitem{Bender2008a}  \href{https://doi.org/10.1103/PhysRevLett.100.110402}{C. M. Bender and P. D. Mannheim, Phys. Rev. Lett. {\bf 100}, 110402 (2008).}

\bibitem{Bender2008b}  \href{https://doi.org/10.1103/PhysRevD.78.025022} {C. M. Bender and P. D. Mannheim, Phys. Rev. D {\bf 78}, 025022 (2008).}

\bibitem{Mannheim2011a} \href{https://doi.org/10.1007/s10714-010-1088-z}{P. D. Mannheim, Gen. Rel. Gravit. {\bf 43}, 703 (2011).}

\bibitem{Mannheim2018} \href{https://doi.org/10.1103/PhysRevD.98.045014}{P. D. Mannheim, Phys. Rev. D {\bf 98}, 045014 (2018).}

\bibitem{Mannheim1992}\href{https://doi.org/10.1086/171358}{P. D. Mannheim, Astrophys. J. \textbf{391}, 429  (1992).}





\bibitem{Riess1998} \href{https://doi.org/10.1086/300499}{A. G. Riess  et. al., Astron. J. \textbf{116}, 1009 (1998).}

\bibitem{Perlmutter1999} \href{https://doi.org/10.1086/307221}{S. Perlmutter  et. al., Ap. J.  \textbf{517}, 565 (1999).}



\bibitem{Mannheim2012a}  \href{https://doi.org/10.1103/PhysRevD.85.124008}{P. D. Mannheim, Phys. Rev. D {\bf 85}, 124008 (2012).}

\bibitem{Amarasinghe2019} \href{https://doi.org/10.1103/PhysRevD.99.083527}{A. Amarasinghe, M. G. Phelps and  P. D. Mannheim, Phys. Rev. D \textbf{99}, 083527 (2019).}

\bibitem{Phelps2019} \href{https://doi.org/10.1007/s10714-020-02757-0} {M.  G. Phelps, A. Amarasinghe  and P. D. Mannheim, Gen. Rel. Gravit. \textbf{52}, 114 (2020).}

\bibitem{Mannheim2020} \href{https://doi.org/10.1103/PhysRevD.102.123535} {P. D. Mannheim, Phys. Rev. D \textbf{102}, 123535 (2020).}


\bibitem{Amarasinghe2020} \href{https://arxiv.org/abs/2011.02440}{A. Amarasinghe and  P. D. Mannheim, \textit{Cosmological Fluctuations on the Light Cone}, arXiv:2011.02440 [gr-qc].}

 \bibitem{Hoyle1964} \href{https://doi.org/10.1098/rspa.1964.0227}{F. Hoyle and J. V. Narlikar, Proc. Roy. Soc. A. {\bf 282}, 191 (1964).}
 
 \bibitem{Stelle1977} \href{https://doi.org/10.1103/PhysRevD.16.953}{K. S. Stelle, Phys. Rev. D {\bf 16}, 953 (1977).}
 
 \bibitem{Stelle1978} \href{https://doi.org/10.1007/BF00760427}{K. S. Stelle, Gen. Rel. Gravit. {\bf 9}, 353 (1978).}
 
 \bibitem{Adler1982} \href{https://doi.org/10.1103/RevModPhys.54.729}{S. L. Adler, Rev. Mod. Phys. {\bf 54}, 729 (1982).}
 
 \bibitem{Lee1982} \href{https://doi.org/10.1103/PhysRevD.26.934}{S.-C. Lee and P. van Nieuwenhuizen, Phys. Rev. D {\bf 26}, 934 (1982).}
 
  \bibitem{Zee1983} \href{https://doi.org/10.1016/0003-4916(83)90286-5}{A. Zee, Ann. Phys. (N. Y.) {\bf 151}, 431 (1983).}
  

  
 \bibitem{Riegert1984a} \href{https://doi.org/10.1016/0375-9601(84)90648-0}{R. J. Riegert, Phys. Lett. A {\bf 105}, 110 (1984).}
  
\bibitem{Riegert1984b} \href{https://doi.org/10.1103/PhysRevLett.53.315}{R. J. Riegert, Phys. Rev. Lett. {\bf 53}, 315 (1984).}

\bibitem{Teyssandier1989} \href{https://doi.org/10.1088/0264-9381/6/2/016}{P. Teyssandier, Class. Quantum Grav. {\bf 6}, 219 (1989).}


\bibitem{'tHooft2010a} \href{https://arxiv.org/abs/1009.0669}{G. 't Hooft,  {\it Probing the small distance structure of canonical quantum gravity using the conformal group}, September 2010. (arXiv:1009.0669 [gr-qc])}

\bibitem{'tHooft2010b}  \href{https://arxiv.org/abs/1011.0061}{G. 't Hooft,  {\it The conformal constraint in canonical quantum gravity}, November 2010. (arXiv:1011.0061 [gr-qc])}


\bibitem{'tHooft2011} \href{https://doi.org/10.1007/s10701-011-9586-8}{G. 't Hooft,  Found. Phys. {\bf 41}, 1829 (2011).}

\bibitem{'tHooft2015a} \href{https://doi.org/10.1142/S0218271815430014}{G. 't Hooft, Int. J. Mod. Phys. D {\bf 24}, 1543001 (2015).} 

\bibitem{Maldacena2011}  \href{https://arxiv.org/abs/1105.5632}{J. Maldacena, {\it Einstein gravity from conformal gravity}, May 2011. (arXiv:1105.5632 [hep-th])}

\bibitem{Horne2016} \href{https://doi.org/10.1093/mnras/stw506}   {K.Horne, Mon. Not. R. Astron. Soc. \textbf{458}, 4122 (2016).}

\bibitem{Lifshitz1946} \href{https://doi.org/10.1007/s10714-016-2165-8}{E. M. Lifshitz, J. Phys. (USSR) {\bf 10}, 116 (1946) (republished as Gen. Rel. Gravit. \textbf{49}, 18  (2017)).}

\bibitem{Bardeen1980} \href{https://doi.org/10.1103/PhysRevD.22.1882}{J. M. Bardeen, Phys. Rev. D \textbf{22}, 1882 (1980).}


\bibitem{Kodama1984} \href{https://doi.org/10.1143/PTPS.78.1}{H. Kodama and M. Sasaki, Prog. Theo. Phys. Suppl. {\bf 78}, 1 (1984).}

 \bibitem{Stewart1990} \href{https://doi.org/10.1088/0264-9381/7/7/013}{J. M. Stewart, Class. Quantum Grav. \textbf{7}, 1169 (1990).}

\bibitem{Mukhanov1992} \href{https://doi.org/10.1016/0370-1573(92)90044-Z} {V. F. Mukhanov, H. A. Feldman and R. H. Brandenberger, Phys. Rept. \textbf{215}, 203 (1992).}
 
\bibitem{Ma1995} \href{https://doi.org/10.1086/176550} {C.-P.  Ma and E. Bertschinger, Astrophys. J. \textbf{455}, 7 (1995).}
    
\bibitem{Bertschinger1996} E. Bertschinger, {\it Cosmological Dynamics}, in Cosmology and Large Scale Structure, proc. Les Houches Summer School, Session LX, ed. R. Schaeffer, J. Silk, M. Spiro and J. Zinn-Justin (Amsterdam: Elsevier Science) (1996).

 \bibitem{Zaldarriaga1998} \href{https://doi.org/10.1086/305223}{M. Zaldarriaga, U.  Seljak and E. Bertschinger, Astrophys. J. \textbf{494}, 491 (1998).}
 
 \bibitem{Weinberg1972} S. Weinberg, {\it Gravitation and Cosmology:
Principles  and Applications of the General Theory of Relativity} (Wiley, New York, 1972).
 
 \bibitem{footnoteZ2} The conformal theory $T^{\mu\nu}$ has to be derived by variation of a conformal invariant action with respect to the metric, and as such it will transform as $T^{\mu\nu}\rightarrow e^{-6\alpha(x)}T^{\mu\nu}$ under $g_{\mu\nu}(x)\rightarrow e^{2\alpha(x)}g_{\mu\nu}(x)$. With $W^{\mu\nu}$ transforming as $W^{\mu\nu}\rightarrow e^{-6\alpha(x)}W^{\mu\nu}$ the gravitational equation of motion $4\alpha_gW^{\mu\nu}=T^{\mu\nu}$ is conformal invariant. Also we should note that there is relation of a conformal $T^{\mu\nu}$ to Sakharov's ideas on induced gravity. Specifically, Sakharov calculated the vacuum expectation value $\langle \Omega|T^{\mu\nu}|\Omega\rangle$ of a general $T^{\mu\nu}$ in the presence of a background metric, and with a cutoff $\Lambda$ found that $\langle \Omega|T^{\mu\nu}|\Omega\rangle=c_1\Lambda^4 g^{\mu\nu}+c_2\Lambda^2(R^{\mu\nu}-(1/2)g^{\mu\nu}R)$ plus higher-order gravity terms. Given (\ref{1.4z}), these higher order terms would begin with ${\rm ln}\Lambda^2(c_3W_{(2)}^{\mu\nu}+c_4W_{(1)}^{\mu\nu})$. (Here all the $c_i$ are finite constants.) Thus in general we induce both the Einstein tensor and a cosmological constant term together with non-leading higher-order gravity terms. However, if we take $T^{\mu\nu}$ to be conformal we would obtain $c_1=0$, $c_2=0$, and  $c_4=-c_3/3$, to then only induce the conformal gravity Bach tensor $W^{\mu\nu}=W_{(2)}^{\mu\nu}-(1/3)W_{(1)}^{\mu\nu}$ that is given in (\ref{1.3z}). Thus it is possible to induce conformal gravity.
 
 
 
\bibitem{Bach1921} \href{https://doi.org/10.1007/BF01378338}{R. Bach, Math. Z.  \textbf{9}, 110 (1921).}

\bibitem{footnoteZ3} Here the $\gamma^c$ form a set of fixed basis Dirac gamma matrices, the $V^{\mu}_c(x)$ are vierbeins, and the spin connection $\Gamma_{\mu}(x)$ is given by $\Gamma_{\mu}(x)=-(1/8)[\gamma_a,\gamma_b](V^b_{\nu}\partial_{\mu}V^{a\nu}+V^b_{\lambda}\Lambda^{\lambda}_{\phantom{\lambda}\nu\mu}V^{a\nu})$ where $\Lambda^{\lambda}_{\phantom{\lambda}\nu\mu}=(1/2)g^{\lambda \sigma}(\partial_{\nu}g_{\mu\sigma} +\partial_{\mu}g_{\nu\sigma}-\partial_{\sigma}g_{\nu\mu})$.


\bibitem{footnoteZ4} Despite the fact that the global cosmological $G_{\rm eff}$ is negative, the parameter $\beta^*$ that appears in the Newtonian  potential term in (\ref{1.16x}) is positive, since the local inhomogeneous gravity associated with a static source is not controlled by the global $G_{\rm eff}$ associated with a homogeneous comoving geometry and a vanishing Weyl tensor but by an induced local $G$ that is associated with an inhomogeneous geometry and a non-vanishing Weyl tensor \cite{Mannheim2006}.

\bibitem{footnoteZ5} In order for $\gamma_0$ to be a static parameter that does not vary from epoch to epoch, it could not be related to the epoch-dependent Hubble parameter, but it could be related to the epoch-independent spatial three-curvature $k$, just as is in fact found to be the case. 

\bibitem{footnoteZ6} Since (\ref{1.20y}) is based on the matter-fluid-independent (\ref{1.15z}), and since (\ref{1.20y}) does fit the redshift less than two accelerating universe data, any conformal model in which the matter sector perfect fluid does contribute would have to be such that it would only be relevant at redshifts higher than a redshift of two or so.


\bibitem{footnoteZ7} From this point we set $c=1$, so $t$ or $\tau$ means $ct$ or $c\tau$.

\bibitem{footnoteZ8} As discussed in \cite{Amarasinghe2019}, even though the gravitational $\delta G_{\mu\nu}$ is not gauge invariant (only the full $\Delta_{\mu\nu}$ as defined in (\ref{2.24y}) is gauge invariant in the matter sector), the purely gravitational $\delta W_{\mu\nu}$ already is gauge invariant on its own.

\bibitem{Mannheim2016} \href{https://doi.org/10.1103/PhysRevD.93.068501}{P. D. Mannheim, Phys. Rev. D \textbf{93}, 068501 (2016).}

\bibitem{footnoteZ9} In practice this is most easily facilitated by working in a gauge in which the $\psi$ fluctuation that appears in $\delta\hat{R}$ and $\delta \hat{P}$ in (\ref{2.30y}) is set to zero, since then $\delta\hat{R}=\delta R$ and $\delta \hat{P}=\delta P$, and $\delta \hat{P}/\delta \hat{R}=\delta p_m/\delta \rho_m$.

\bibitem{footnoteZ10} In \cite{Phelps2019} it is shown that for $k<0$ there are no solutions to $(\tilde{\nabla}_a\tilde{\nabla}^a+3k)S=0$ where $S$ is a scalar that are well behaved at both $\chi =\infty$ and $\chi=0$. Thus we can factor out the overall $\tilde{\nabla}_a\tilde{\nabla}^a+3k$ factor on the right-hand side of  (\ref{4.2x}).


\bibitem{footnoteZ11} Since $\eta=-24\alpha_g/S_0^2$ and $\sigma=-2\lambda S_0^2$, $\kappa=\sigma\eta$ is given by $\kappa=48\lambda \alpha_g$. The dimensionless $\kappa$ is thus proportional to the product of the gravitational and the scalar field coupling constants, while having no explicit dependence on $S_0$. 


\bibitem{Bander1966} \href{https://doi.org/10.1103/RevModPhys.38.346} {M. Bander and C. Itzykson, Rev. Mod. Phys. \textbf{38}, 346 (1966).}

\bibitem{footnoteZ12} For non-small comoving times we note that from the power series expansion of $F(-\zeta,\zeta+1;1-\mu;(1-z)/2)$ it follows that $P^{0}_{\zeta}(z)$ is real if  $z$ is real and $\zeta$ is in the conical function form $\zeta=-1/2+i\lambda$ where $\lambda$ is real. Also, $P^{0}_{-1/2+i\lambda}(z)$ with real $z$ has an infinite number of real zeros when $z>1$ , i.e., for all $\xi$ since $z=\cosh\xi$, with oscillations that are damped. Moreover, numerically we have found in some typical cases that ${\rm RE}[P^{i\nu}_{-1/2+i\lambda}(z)]$ with real nonzero $\nu$ and real $z$ also has oscillations that are damped. For confirmation, we note that at large $z$ $P^{i\nu}_{\zeta}(z)$ is known to behave as $z^{\zeta}2^{\zeta}\Gamma[\zeta+1/2]/(\Gamma[1/2]\Gamma[\zeta -i\nu+1])$ for any $\zeta$ and $\nu$. Thus with $\zeta=-1/2\pm (1/4-1/\kappa)^{1/2}$ as given  (\ref{4.19x}), (\ref{5.11y}), (\ref{6.29y}), the scalar, vector and tensor comoving time solutions will only oscillate at late times if $1/\kappa>1/4$, i.e., if  $0<\kappa<4$. Since $\kappa=\sigma\eta$ and since $\sigma$ is positive, this can only occur if $\eta$ is positive, just as we noted is in fact the case \cite{Mannheim2011a,Mannheim2016}. And with $\kappa<4$ the coupling constants are constrained to obey $\lambda\alpha_g<12$. (This result replaces the result presented in \cite{Mannheim2020}, where it was thought that one gets oscillating solutions for any positive $\eta$. )
























\end{thebibliography}
\end{document}